\documentclass[11pt,a4paper]{article}
\usepackage{jheppub}
\usepackage{bm}
\usepackage{amsmath}	

\usepackage{lineno,hyperref}
\usepackage{amsmath}
\usepackage{amssymb}
\modulolinenumbers[5]

\def\MSbar{\ensuremath{\overline{\mathrm{MS}}}}
\def\DIANA{\texttt{DIANA}}
\def\MATAD{\texttt{MATAD} }
\def\FORM{\texttt{FORM} }
\def\SARAH{\texttt{SARAH} }
\def\PyRATE{\texttt{PyR@TE}}
\def\hc{\ensuremath{\mathrm{h.c.}}}

\newcommand\tr[1]{\ensuremath{\mathrm{tr}#1}}
\def\myRe{\ensuremath{\mathrm{Re}}}
\def\myIm{\ensuremath{\mathrm{Im}}}
\def\lamS{\ensuremath{\Lambda_{00}}}
\def\mS{\ensuremath{M_0}}
\def\lamV{\ensuremath{\vec{\Lambda}}}
\def\mV{\ensuremath{\vec{M}}}
\def\lamT{\ensuremath{{\Lambda}}}
\newcommand\trlam[2]{\ensuremath{\mathrm{tr}^{#1}\Lambda^{#2}}}
\def\LambdaMuNu{\ensuremath{\Lambda_{\mu\nu}}}
\def\mMu{\ensuremath{M_\mu}}

\title{On three-loop RGE for the Higgs Sector of 2HDM}

\author{A.V. Bednyakov}
\emailAdd{alexander.bednyakov@jinr.ru}

\affiliation{
	Joint Institute for Nuclear Research, \\ 
	Joliot-Curie, 6, Dubna  141980, Russia}

\affiliation{Dubna State University,\\ 
	     Universitetskaya, 19, Dubna 141982, Russia}

\affiliation{P.N. Lebedev Physical Institute of the Russian Academy of Sciences, \\
	  Leninskii pr., 5, Moscow 119991, Russia}

\abstract{
	We discuss renormalization group equations (RGE) for the parameters of the Higgs sector in general Two-Higgs-Doublet Model (2HDM).
	We present the three-loop results but consider only contributions due to self-couplings of the Higgs doublets. We study the structure of RGE and express beta-functions in terms of reparametrization invariants with respect to higgs-basis rotations. The Cayley-Hamilton theorem is utilized to reduce both the number of independent tensor structures in matrix RGE and the number of invariants to a minimal set. As a by-product of our calculation we discovered that two-loop RGE of the scalar sector in general QFT with multiple higgses were not properly implemented in a number of public packages. The latter give rise to a wrong result when mixing in the scalar sector is allowed. 
}

\arxivnumber{1809.04527}

\begin{document}
\maketitle

\section{\label{sec:intro}Introduction}

	The Standard Model (SM) was established in mid-1970s. Its success is incredible: even after almost half a century, no significant deviations from the SM predictions were found.  %remains the only model that
	Given a minimal set of parameters, the SM provides a very precise description of different phenomena in Modern Particle Physics. 
	To confront its predictions with ongoing and future experiments, one is forced to take various radiative corrections into account and, in many cases, carry out certain kind of re-summation. 
	A convenient tool to deal with high-order terms in perturbative expansion is dimensional regularization \cite{tHooft:1972tcz} 
	accompanied by modified minimal ($\MSbar$) subtractions  of infinities. The latter appear in loop integrals and manifest itself as poles in $\epsilon = (4-d)/2$.

	In \MSbar-renormalization scheme the model parameters depend on auxiliary scale $\mu$ and their numerical values at different scales are related by differential renormalization group equations (RGE). While boundary conditions should be extracted from experiment, 
	the RG functions (beta functions and anomalous dimensions) can be calculated order-by-order in perturbation theory.
	Solution of RGE allows one to improve the precision of finite-order predictions by re-summing certain logarithmic corrections into redefinition of the model parameters. 

For the parameters of the SM Lagrangain  three-loop RG functions are known from recent 
computations: the gauge coupling are considered in refs.~\cite{Mihaila:2012fm,Mihaila:2012pz,Bednyakov:2012rb}, 
the results for Yukawa couplings can be found in refs.~\cite{Bednyakov:2012en,Bednyakov:2013cpa,Bednyakov:2014pia}, and
refs.~\cite{Chetyrkin:2013wya,Bednyakov:2013eba} are devoted to the SM Higgs-potential parameters.
There are also partial four-loop results available in literature  (see., refs~\cite{Martin:2015eia,Bednyakov:2015ooa,Zoller:2015tha,Chetyrkin:2016ruf}. Recently, five-loop RG functions in pure QCD have been calculated~\cite{Baikov:2016tgj,Herzog:2017ohr,Luthe:2017ttg}.
Among other things, all the results were immediately applied to state-of-the-art studies \cite{Buttazzo:2013uya,Bednyakov:2015sca} of the vacuum-stability problem in the SM.

In spite of the above-mentioned success of the SM, there are well-known issues (related to dark matter, fine-tuning, etc.)  that prevent us from treating the SM as the most fundamental theory of particle interactions (see, e.g., ref.~\cite{Kazakov:2018yax}). 
Among different possibilities to go beyond the SM (BSM) one can consider an extension with an additional Higgs doublet - the so-called 
Two-Higgs-Doublet Model (2HDM) (for review see refs.~\cite{Branco:2011iw,Ivanov:2017dad}). 
The model % being one of the simplest (yet renormalizable) alternatives of the SM 
predicts new scalar states in the spectrum --- two neutral $H,A$ and one charged $H^\pm$ higgs bosons. 
Being (linear combinations of) components of the $\mathrm{SU}(2)$ doublets, their interactions with vector fields are fixed by postulated gauge symmetry, 
but there is a freedom in self-interactions and fermion Yukawa couplings.

Recently, three-loop beta-functions for the gauge and Yukawa sector of general (Type-III) 2HDM were found in ref.~\cite{Herren:2017uxn}. In this paper, we continue the study of the RG functions in 2HDM and calculate certain three-loop contributions to the beta functions of the higgs self couplings and anomalous dimensions of the higgs mass parameters entering general Higgs potential. 

We restrict ourselves to the corrections due to the scalar self-interactions only and for the moment we neglect both gauge and Yukawa couplings of the Higgs fields. Nevertheless, we consider different parameterization of the scalar sector. In addition, we compute the scale dependence of reparametrization invariants (see, e.g, ref.~\cite{Ivanov:2005hg}), which are constructed from the Higgs potential parameters, but contrary to the latter, do not depend on the choice of Higgs basis. 
For convenience, all the RG functions considered in this paper\footnote{
Up to the three-loop order.} are available as ancillary files of the arXiv version of the paper.

It is worth mentioning that we have tried to compare the two-loop beta-functions obtained by direct calculations with the RG-functions extracted from the well-known results for a general renormalizable QFT model~\cite{Machacek:1984zw,Luo:2002ti}. We have found that application of the available general result to the case with many scalar fields requires some care. We discovered that, e.g., current versions of \SARAH \cite{Staub:2013tta} and \PyRATE~\cite{Lyonnet:2016xiz}, when applied to the case of Type-III 2HDM, give rise to a wrong result at two loops (see Section~\ref{sec:rge2_lambda_m} for details).

The paper is organized as follows. In section \ref{sec:potential} we introduce the 2HDM Higgs potential and discuss various parametrizations of the Higgs sector.
In section~\ref{sec:RG_structure} the structure of the RG functions in one particular parametrization, which involve scalars $\lamS$, $\mS$, 3-vectors $\lamV$, $\mV$ and a symmetric $3\times3$ tensor $\lamT$, is elaborated.   
section \ref{sec:RG_procedure} is devoted to the description of the renormalization procedure for the above-mentioned quantities. The corresponding three-loop RG functions can be found in section~\ref{sec:res_3loop}. We discuss subtleties in the interpretation of the well-known two-loop expressions \cite{Machacek:1984zw,Luo:2002ti} and present our results for self-couplings $\lambda_i$ and masses $m^2_{ij}$ in section~\ref{sec:rge2_lambda_m}. Our conclusions can be found in section~\ref{sec:conclusios}. In a series of appendices we provide some details on the reparametrization-invariant counting via Hilbert Series (\ref{app:Hilbert_series}) and present useful identities for $\lamT$ (\ref{app:identities}). In addition, the RG-functions of the reparametrization invariants are  given in appendix~\ref{app:invariants_rge}.

	\section{\label{sec:potential}The scalar potential of 2HDM}
	The most general renormalizable Higgs potential can be written in the following form
\begin{align}
	V_H & = 
		  m_{11}^2 \Phi_1^\dagger \Phi_1
		+ m_{22}^2 \Phi_2^\dagger \Phi_2
		- \left(m_{12}^2 \Phi_1^\dagger \Phi_2 + \hc\right)
		\nonumber\\
		& 
		+ \frac{1}{2} \lambda_1 \left(\Phi_1^\dagger \Phi_1\right)^2 
		+ \frac{1}{2} \lambda_2 \left(\Phi_2^\dagger \Phi_2\right)^2 
		+  \lambda_3 \left(\Phi_1^\dagger \Phi_1\right)\left(\Phi_2^\dagger \Phi_2\right) 
		+  \lambda_4 \left(\Phi_1^\dagger \Phi_2\right)\left(\Phi_2^\dagger \Phi_1\right) 
		\nonumber\\
		&+ \left[ 
			\frac{1}{2} \lambda_5 \left(\Phi_1^\dagger \Phi_2\right)^2
			+ \lambda_6 \left(\Phi_1^\dagger \Phi_1\right) \left(\Phi_1^\dagger \Phi_2\right)
			+ \lambda_7 \left(\Phi_2^\dagger \Phi_2\right) \left(\Phi_1^\dagger \Phi_2\right)
			+ \hc
		\right]
\label{eq:V_gen}
\end{align}
with $\Phi_{1,2}$ being $SU(2)$ doublets. The self-couplings $\lambda_{1-4}$ and the mass parameters  $m_{11}^2$, $m_{22}^2$ are real, while $\lambda_{5-7}$, and $m_{12}^2$ can be complex. Not all of these fourteen (real) parameters are physical due to the freedom in redefinition of Higgs basis by a unitary rotation
\begin{align}
	\Phi_a \to U_{ab} \Phi_b, \quad U \in \mathrm{U}(2),
\end{align}
where $a,b=1,2$ enumerate the doublets. It is easy to see that the overall phase of $U$ does not impact the change for the couplings and masses so in what follows we restrict ourselves to $U\in \mathrm{SU}(2)$. The three parameters of $\mathrm{SU}(2)$ rotation can be used to get rid of three out of 14 parameters of the potential and, thus, we are left only with 11 independent quantities. 

There is an alternative notation \cite{Botella:1994cs}
\begin{align}
	V_H & = \frac{1}{2} \lambda_{ab,cd} (\Phi^\dagger_a \Phi_b)(\Phi^\dagger_c \Phi_d) 
	+ m^2_{ab} (\Phi^\dagger_a \Phi_b),\quad \lambda_{ab,cd} = \lambda_{cd,ba}, \quad
	m^2_{ba} = m^{\dagger2}_{ab},
\label{eq:V_m_lambda}
\end{align}
which can be used as an intermediate step to rewrite the self-couplings (see refs.~\cite{Branco:2011iw,Ivanov:2017dad}) 
\begin{align}
	\lambda_{ab,cd} & = \frac{1}{2} \Lambda_{\mu\nu} \sigma^\mu_{ab} \sigma^\nu_{cd} 
	= \frac{1}{2} \left[
		\lamS \delta_{ab} \delta_{cd} + \lamV \left( \vec{\sigma}_{ab} \delta_{cd}
		+ \delta_{ab} \vec{\sigma}_{cd} \right)
		+  \vec{\sigma}_{ab} \cdot  \Lambda_{ij} \cdot \vec{\sigma}_{cd} 
	\right],
	\label{eq:lambda_to_Lambda}
\end{align}
\begin{align}
	\Lambda_{\mu\nu} & = \frac{1}{2} \lambda_{ab,cd} \sigma_\mu^{ba} \sigma_\nu^{dc}
	= 
	\begin{pmatrix}
		\frac{\lambda_1 + \lambda_2}{2} + \lambda_3 & 
	 \myRe\left(\lambda_6 + \lambda_7\right) & 
	-\myIm\left(\lambda_6 + \lambda_7\right) &
		\frac{\lambda_1 - \lambda_2}{2} \\
		\phantom{-}\myRe\left(\lambda_6 + \lambda_7\right) &
		\lambda_4 + \myRe \left( \lambda_5\right) &
		- \myIm \left( \lambda_5 \right) &
		\phantom{-}\myRe\left(\lambda_6 - \lambda_7\right)
	\\
	- \myIm \left( \lambda_6 + \lambda_7\right)&
	- \myIm \left( \lambda_5 \right) &
	\lambda_4 - \myRe \left( \lambda_5\right) &
	- \myIm \left( \lambda_6 - \lambda_7\right) \\
	\frac{\lambda_1 - \lambda_2}{2} &
	\myRe\left( \lambda_6 - \lambda_7\right) &
	- \myIm \left( \lambda_6 - \lambda_7 \right) &
	\frac{\lambda_1 + \lambda_2}{2} - \lambda_3
	\end{pmatrix}
\label{eq:LambdaMuNudef}
\end{align}
in terms of a scalar $\lamS$, a vector $\lamV$ and a symmetric matrix $\lamT$, where $\sigma^\mu \equiv (1,\vec{\sigma})$, 
	$\mu,\nu=0,1,2,3$, $i,j=1,2,3$ and the euclidean metric is used both for four- and three-dimensional indices. 
The same trick can be used for the mass term:
\begin{align}
	m^2_{ab} & = \frac{1}{2} M_\mu \sigma^\mu_{ab} = \frac{1}{2} \left[ M_0 \delta_{ab} + \vec{M} \vec{\sigma}_{ab} \right],
	\quad
\label{eq:m_to_M} 
\end{align}
% checked the signs of m12
\begin{align}
\mS & = \tr\left[m^2\right] = m^2_{11} + m_{22}^2, 
\quad \mV = \tr \left[m^2 \vec{\sigma} \right] = \left( 
		- 2\myRe\, m_{12}^2,
		  2\myIm\, m_{12}^2,
	  m_{11}^2 - m_{22}^2
  \right).
\end{align}
We can also decompose the tensor
\begin{align}
	\Phi_a \Phi^\dagger_b & = \frac{1}{2} \left(\Phi^\dagger \Phi\right) \delta_{ab}
	+ \frac{1}{2} \left(\Phi^\dagger \sigma^n \Phi\right) \sigma^n_{ab}
	= r_0 \, \delta_{ab} + \vec{r} \cdot \vec{\sigma}_{ab} = r_\mu \sigma^\mu_{ab}
	\label{eq:PhiPhi_decomposition}
\end{align}
in terms of a singlet $r_0$ and a vector $\vec{r}$.  
By means of eqs.\eqref{eq:lambda_to_Lambda},\eqref{eq:m_to_M}, and \eqref{eq:PhiPhi_decomposition} one can rewrite the potential \eqref{eq:V_m_lambda} as 
\begin{align}
	V_H = \frac{1}{4} \Lambda_{\mu\nu} 
	  r_\rho 
	 r_{\sigma} 
	 \left[
	\sigma^\mu_{ab} \sigma^\rho_{ba}
	\sigma^\nu_{cd} \sigma^\sigma_{dc}
\right]
	+ \frac{1}{2} M_\mu r_\nu 
	\left[
	   \sigma^\mu_{ab}
	   \sigma^\nu_{ba}
   	\right]
	=
	M_\mu r^\mu + \Lambda_{\mu\nu} r^\mu r^\nu
	\label{eq:V_M_Lambda}
\end{align}

Under a Higgs-basis change $\Phi_a \to U_{ab} \Phi_b$, $U_{ab} \in \mathrm{SU}(2)$, 
$\lamS$ and $\mS$ transform as singlets, while $\lamV$ and $\mV$ transform as triplets under the corresponding $\mathrm{SO}(3)$ rotation
\begin{align}
	R_{ij}(U) &  = \frac{1}{2} \tr\left[ U^\dagger \sigma_i U \sigma_j \right]. 
\end{align}
The symmetric $3\times3$ matrix $\lamT\equiv\{\Lambda_{ij}\}$ can be decomposed\footnote{In what follows, we do not use this decomposition.} into a singlet $\trlam{}{}$ and
a five-plet $\tilde \Lambda_{ij} \equiv \left[\Lambda_{ij} - \frac{1}{3} \trlam{}{} \delta_{ij}\right]$. 

\section{\label{sec:RG_structure}The structure of RG functions}
	The parametrization of the quartic couplings in terms of $\lamS$, $\lamV$ and $\lamT$ turns out to be very convenient for calculation of RGE in the scalar sector. The main advantage of the approach is that we need to deal with at most two indices instead of four.  
	In addition, the Cayley-Hamilton theorem, which states that the square  $3\times 3$ matrix $\lamT$ satisfies its own characteristic equation 
	\begin{align}
	\lamT^3 & = 
		\trlam{}{} \lamT^2 
		- \frac{1}{2} \left( \trlam{2}{} - \trlam{}{2} \right)\lamT 
		+ \frac{1}{3!} \left( 
			\trlam{3}{} - 3 \trlam{}{} \trlam{}{2} + 2 \trlam{}{3}
		\right),
	\label{eq:CayleyHamilton}
\end{align}
can be used to get rid of high powers $\Lambda^n$ ($n\geq 3$) appearing at the intermediate steps of calculation.

Due to eq.~\eqref{eq:CayleyHamilton} we can enumerate possible structures that can appear in beta-functions for the components of $\LambdaMuNu$ %$\lamS$, $\lamV$ and $\lamT$ 
($t=\ln \mu^2$, $h=(16 \pi^2)^{-1}$):
\begin{align}
	\frac{d \LambdaMuNu}{d t} & =   \beta_{ \LambdaMuNu}  = 
	\sum_{l=1}^{\infty} h^l \beta^{(l)}_{\LambdaMuNu}, %\qquad \LambdaMuNu = \left\{ \lamS, \lamV, \lamT \right\}.
\label{eq:blam_def}
\end{align}

Since $\lamS$ is an invariant (singlet w.r.t higgs-basis transformations), only reparametrization invariants can enter $\beta_{\lamS}$.  Given $\lamS$, $\lamV$, and $\lamT$ one can introduce the following independent invariants\footnote{All other scalars of the form $\lamV \cdot \lamT^n \cdot \lamV$ $(n\geq 3)$ and $\trlam{}{m}$ $(m>3)$ can be reduced to \eqref{eq:quartic_invariants} via \eqref{eq:CayleyHamilton}.} $I_{i,j}$ (c.f. 
\cite{Ivanov:2005hg}): 
\begin{subequations}
\begin{align}
	I_{1,1} & = \lamS, \hspace{1.5cm}
I_{1,2}  = \trlam{}{},  
\label{eq:qinv_1}\\ 
I_{2,1} & =  \lamV \cdot \lamV, \hspace{1.3cm}
I_{2,2}   = \trlam{}{2}, 
\label{eq:qinv_2}\\ 
I_{3,1} & = \lamV \cdot \lamT \cdot \lamV,  \hspace{0.8cm}
I_{3,2}  = \trlam{}{3},  
\label{eq:qinv_3}\\ 
I_{4,1} & = \lamV \cdot \lamT^2 \cdot \lamV. 
\label{eq:qinv_4}
\end{align}
\label{eq:quartic_invariants}
\end{subequations}
The first index in $I_{i,j}$ corresponds to the order (or degree) of the invariant, i.e., the total power of the $\Lambda_{\mu\nu}$ components entering $I_{i,j}$.  There is also an invariant of order six  
\begin{align}
	I_{6,1}& = \lamV \cdot\left[(\lamT \cdot \lamV) \times (\lamT^2 \cdot \lamV)\right],
\label{eq:quartic_invariants_six}
\end{align}
which is related (up to a sign) to those presented in eq.~\eqref{eq:quartic_invariants}. A convenient tool to enumerate the invariants is the so-called Hilbert Series (see appendix~\ref{app:Hilbert_series} and references therein).

The beta-function $\beta_{\lamV}$ can be cast into the general form\footnote{We assume that all three vectors in \eqref{eq:beta_LV_gen} are independent and form a basis in 3d, so, e.g., $ \left[ (\lamT \cdot \lamV) \times (\lamT^2 \cdot \lamV) \right]$ does not appear in $\beta_{\lamV}$.}
\begin{align}
	\beta_{\lamV} = a_{0}\, \lamV + a_{1}\, \lamT \cdot \lamV + a_{2} \, \lamT^2 \cdot \lamV 
\label{eq:beta_LV_gen}
\end{align}
with $a_{i}$ being polynomials in invariants \eqref{eq:quartic_invariants}. The beta-function $\beta_{\lamT}$ looks like 
	\begin{align}
	\beta_{\lamT}  =   
	b_{0}\, +
	b_{1}\, \lamT + b_{2} \, \lamT^2 + b_3 \, \lamV \otimes \lamV
	+ b_4 \left( \lamT \cdot \lamV \otimes \lamV +   \lamV \otimes \lamT \cdot \lamV \right)
	%+ b_5 \left( \lamT^2 \cdot \lamV \otimes \lamV +   \lamV \otimes \lamT^2 \cdot \lamV \right),
	+ b_5 \lamT \cdot \lamV \otimes \lamT \cdot \lamV, 
\label{eq:betb_LT_gen}
	\end{align}
	where $\lamV \otimes \lamV\equiv \lamV_{i} \lamV_{j}$, etc. and $b_i$ are again expressed in terms of invariants. By means of Cayley-Hamilton theorem one can also show that other symmetric tensors constructed from $\lamT$ and $\lamV$ are not independent (see appendix~\ref{app:identities} for details). 

	Since in the $\MSbar$ scheme counter-terms are polynomial in momenta and masses \cite{tHooft:1972tcz}, it is clear from dimensional analysis that RGE for $\mMu$ can only involve first powers of the latter. As a consequence,  scalars (invariants) and vectors involving high powers of $\mV$ and $\mS$ will not contribute to the %  in order to write down an ansatz for 
	mass anomalous dimensions, which we define here as
\begin{align}
	\frac{d \mMu}{d t} & = \gamma_{\mMu} =    
	\sum_{l=1}^\infty h^l \gamma_{\mMu}^{(l)}. %\qquad \mMu = \left\{ \mS, \mV \right\}.
	\label{eq:gamma_M_def} 
\end{align}
	Indeed, the RG equation for $\mS$ should be a linear combination of the following reparametrization invariants

\begin{align}
	I_{0,1}  =   \mS,  \quad
	I_{1,3}  = \lamV \cdot \mV,  \quad
	I_{2,3}  = \lamV \cdot \lamT \cdot \mV, \quad
	I_{3,3}  = \lamV \cdot \lamT^2 \cdot \mV 
\label{eq:additional_invariants}
\end{align}
with coefficients being polynomials in the invariants built from $\Lambda_{\mu\nu}$ only. % \eqref{eq:quartic_invariants}.%iging p 
The anomalous dimension $\gamma_{\mV}$ must be a linear combination of the vectors 
\begin{align}
	& \phantom{I_M} \mV, 
	\quad \phantom{I_M} \lamT  \cdot \mV,	
	\quad \phantom{I_M} \lamT^2 \cdot \mV, \\
	&  \,I_M \,\lamV, 
	\quad I_M \,\lamT \cdot \lamV,	
	\quad I_M \,\lamT^2 \cdot \lamV,
\end{align}
where $I_M$ denotes one of the invariants from eq.~\eqref{eq:additional_invariants}. The results of direct evaluation  of Feynman graphs (see, e.g.,  eqs.~\eqref{eq:bmS}, and \eqref{eq:bmV}) confirm this structure. 

\section{\label{sec:RG_procedure}Renormalization procedure}
In order to find RGE for dimensionless couplings we generate diagrams (self-energies $\Gamma_{a}^{b}$, and four-point functions $\Gamma_{ac}^{bd}$) with external $\Phi_a$,$\Phi^\dagger_b$, etc., but rewrite the quartic vertex in terms of $\Lambda_{\mu\nu}$ by means of eq.~\eqref{eq:lambda_to_Lambda}.
	We heavily rely on automatic index-summation algorithms of \FORM \cite{Vermaseren:2000nd,Kuipers:2012rf} to deal with indices of different dimensions in diagrams generated by \DIANA  \cite{Tentyukov:1999is}. 
	To extract the corrections to $\lamS$, $\lamV$, and $\lamT$ from the considered Green functions we apply projectors, which imply summation over external higgs indices. The form of  the projectors can be deduced from eqs.~\eqref{eq:lambda_to_Lambda} and \eqref{eq:PhiPhi_decomposition}. 

	Let us briefly discuss counter-terms originating from the Lagrangian in the notation of eq.~\eqref{eq:V_M_Lambda}. It is convenient to consider additive renormalization of the parameters, i.e.,
\begin{align}
	\mu^{-2\epsilon} \left(\LambdaMuNu\right)_{bare} & = \LambdaMuNu + \delta \LambdaMuNu \\ 
	\left(\mMu\right)_{bare} & = \mMu + \delta \mMu.
\label{eq:gen_ren_params}
\end{align}
The bare bilinear $\Phi_a$ combinations $(r_\mu)_{bare}$ are given by
\begin{align}
	\frac{1}{2}\left( \Phi^\dagger \Phi \right)_{bare} &  \equiv \left(r_0\right)_{bare}  = (z_0^2 + \vec{z}^2) r_0
	+ 2 z_0 \,\vec{z} \cdot \vec{r}, 
\label{eq:PhiPhi_scalar_ren}\\
	\frac{1}{2}\left( \Phi^\dagger \vec{\sigma} \, \Phi \right)_{bare} & \equiv \left(\vec{r}\right)_{bare}\,\,  = (z_0^2 - \vec{z}^2) \vec{r}
	+ 2 \left( z_0 r_0 + \vec{z} \cdot \vec{r}\right) \vec{z} ,
\label{eq:PhiPhi_vector_ren}
\end{align}
where $z_0$ and $\vec{z}$ come from the decomposition of the hermitian field renormalization constant $Z_\Phi$ entering 
\begin{align}
	\left(\Phi_a\right)_{bare} & = (Z_\Phi)_{ab} \Phi_b = \left(z_0 \delta_{ab} + \vec{z} \cdot \vec{\sigma}_{ab} \right) \Phi_b.
\label{eq:Phi_renorm}
\end{align}
The counter-term Lagrangian is obtained by expressing the bare fields and parameters in terms of renormalized ones by means of the above-mentioned equations. % \eqref{eq:PhiPhi_scalar_ren} 	
The expressions for $\delta \LambdaMuNu = \mathcal{O}(\LambdaMuNu^2)$, 
$\delta \mMu = \mathcal{O}(\LambdaMuNu^2)$, $z_0 = 1 + \mathcal{O}(\LambdaMuNu^2)$, and $\vec{z} = \mathcal{O}(\LambdaMuNu^2)$ are determined order by order in perturbation theory. 

The renormalization constants in the $\MSbar$ scheme are extracted from divergent terms of the corresponding loop integrals. Due to this, we made use of the well-known infrared rearrangement (IRR) tricks\cite{Vladimirov:1979zm}, which allow us to modify the infrared structure\footnote{Strictly speaking, this is only possible for logarithmically divergent integrals. However, we can differentiate w.r.t (equivalently, expand in) external momenta and masses to use the trick.}  of the considered integrals and convert them to fully massive bubbles. A modern version\footnote{Available, at \url{https://github.com/apik/matad-ng}.} of the \MATAD \cite{Steinhauser:2000ry} package written in \FORM was used to compute the vacuum integrals. % with exact $d$ dependence.

Given $\delta \LambdaMuNu$ and $\delta \mMu$ we find beta-functions and mass anomalous dimensions via differentiation of the bare parameters \eqref{eq:gen_ren_params} w.r.t. the scale $t = \ln\mu^2$:
\begin{align}
	\beta_{\LambdaMuNu} & = - \epsilon \LambdaMuNu - \left( \epsilon + \frac{d}{d t} \right) \delta \LambdaMuNu, \quad \gamma_{\mMu}  = - \frac{d}{d t} \delta \mMu. 
\label{eq:RG_from_CT}
\end{align}
Both $\delta \LambdaMuNu$ and $\delta\mMu$ involve higher poles in $\epsilon$. However, the corresponding contribution to the RG functions
is canceled due to the so-called pole equations \cite{tHooft:1973mfk}. As a consequence, the finiteness of \eqref{eq:RG_from_CT} in the limit $\epsilon\to0$ serves as a cross-check of the correctness of our final results. It turns out that one needs to utilize various relations (see appendix~\ref{app:identities}) to prove that the pole equations are satisfied.

In order to find $\delta \mMu$ in the $\MSbar$ scheme, it is sufficient to treat the mass term %$\delta \mathcal{L}  = - M_\mu r^\mu$ 
as a perturbation to the massless theory. The corresponding (bare) Lagrangian can be rewritten as 
\begin{align}
	- \delta \mathcal{L}_2 = (M_\mu)_{bare}  \cdot  (\Phi^\dagger \sigma^\mu \Phi )_{bare} = 
	(Z_{\mu\nu} M^\nu) \cdot (\Phi^\dagger \sigma^\mu \Phi )_{bare}  
	=
	M^\nu \left[ \Phi^\dagger \sigma_\nu \Phi \right], 
	\label{eq:mass_operator_ct_relation}
\end{align}
where renormalized operators $[\Phi^\dagger \sigma^\mu \Phi]$ are related to the bare bilinears via
\begin{align}
	\left[ \Phi^\dagger \Phi \right] & =  Z_{00} \left( \Phi^\dagger \Phi \right)_{bare} 
	+ Z_i \left( \Phi^\dagger \sigma^i \Phi \right)_{bare}, \\
	\left[ \Phi^\dagger \sigma_i \Phi\right] & = \tilde Z_i \left(\Phi^\dagger \Phi\right)_{bare} + 
	\left(\Phi^\dagger \sigma_j \Phi\right)_{bare}
	Z_{ji}.
\label{eq:mass_operators}
\end{align}
From \eqref{eq:mass_operator_ct_relation} one can see that the renormalization constants $Z_{00}$, $Z_i$, $\tilde Z_i$ and $Z_{ij}$ also enter mass-parameter renormalization \eqref{eq:gen_ren_params} 
\begin{align}
	(M_0)_{bare} & = Z_{00} M_0 + \tilde Z_i M_i  = M_0 + \delta M_0  = M_0 + 
	\underbrace{(Z_{00} - 1) M_0 + \tilde Z_i M_i}_{\delta M_0},	\\ 
	(M_i)_{bare} & = Z_i M_0  + Z_{ij} M_j 
	= M_i + \delta M_i = M_i + \underbrace{
	Z_i M_0  + (Z_{ij }- \delta_{ij}) M_j}_{\delta M_i}. 
\end{align}
Due to this, we extract the mass-parameter counter-terms not from massive self-energies with external $\Phi^\dagger$ and $\Phi$, but from divergences of auxiliary three-point functions with an additional $(\Phi^\dagger \sigma^\mu \Phi)$-operator insertion at zero momentum. The latter are computed by means of the above-mentioned IRR trick.  

\section{\label{sec:res_3loop}Three-loop RGE for $\Lambda_{\mu\nu}$ and $M_\mu$}

The procedure discussed in the previous section was used to find RG functions for the Higgs potential parameters~\eqref{eq:V_M_Lambda}. % were obtained along the lines presented earlier. 
The one-, two- and  three-loop results for $\LambdaMuNu$ %$\lamS$, $\lamV$ and $\lamT$ 
are  given by the expressions:
{\allowdisplaybreaks 
\begin{subequations}
\begin{align}
\beta^{(1)}_{\lamS} & = 4\lamS^2 + 6 \vec{\Lambda}^2  +  
		\trlam{}{2} + \trlam{}{} \cdot \lamS, \label{eq:blamS1l} \\
\beta^{(2)}_{\lamS} & = - 8 \trlam{}{3} + \trlam{}{} \cdot \trlam{}{2}  
			+ 0 \cdot \trlam{3}{}
			- \frac{57}{4} \lamS^3 - \frac{11}{2} \lamS^2 \cdot \trlam{}{}
			\nonumber\\
			& + \frac{5}{4} \lamS \cdot \trlam{2}{}
			- \frac{27}{2} \lamS \cdot \trlam{}{2}
			- 66 \lamS \cdot \lamV^2 - 2 \trlam{}{} \cdot \lamV^2
			- 49 (\lamV \cdot \lamT \cdot \lamV)
			,\label{eq:blamS2l} \\
\beta^{(3)}_{\lamS} & = 
\lamS^4 \left(  \frac{389}{4}  + \frac{93}{2} \zeta_3\right)
		  -  \trlam{4}{} \left( 2 - \frac{9}{2} \zeta_3\right) 
		  + \lamS^3 \cdot \trlam{}{} \left( \frac{975}{16} + 24 \zeta_3 \right)
		  + \frac{51}{16} \lamS \cdot \trlam{3}{} 
		  \nonumber \\
		  & 
		  - \lamS^2 \cdot \trlam{2}{} \left(\frac{25}{2} + 9 \zeta_3 \right)
		  + \lamS^2 \cdot \trlam{}{2} \left( \frac{2729}{16} + 90 \zeta_3 \right)
		  + \lamS^2 \cdot \lamV^2 \left( \frac{7191}{8} + 432 \zeta_3 \right) 
		  \nonumber \\
		  & 
		  + \trlam{2}{} \cdot \trlam{}{2} \left( \frac{157}{16} - 24 \zeta_3 \right) 
		  + \trlam{2}{} \cdot \lamV^2 \left( \frac{23}{8} - 6 \zeta_3 \right)
		  - \lamS \cdot \trlam{}{} \cdot \trlam{}{2}
		  \left( \frac{279}{8} + 12 \zeta_3 \right)
		  \nonumber\\
		  & 
		  + \lamS \cdot \trlam{}{} \cdot \lamV^2 \left( 83 + 24 \zeta_3 \right)
		  + \lamS (\lamV \cdot \Lambda \cdot \lamV)
		  \left( 1195 + 624 \zeta_3 \right)
		  + \lamS \cdot \trlam{}{3} \left( 169 + 96 \zeta_3 \right)
		  \nonumber \\
		  & 
		  + \trlam{}{} \cdot ( \lamV \cdot \Lambda \cdot \lamV)
		  \left(38 - 12 \zeta_3 \right)
		  + \trlam{}{} \cdot \trlam{}{3} 
		  \left(20 + 36 \zeta_3 \right)
		  - \frac{263}{8} \trlam{2}{2} 
		  \nonumber\\
		  & 
		  + \trlam{}{2} \cdot \lamV^2 \left(\frac{57}{4} + 54 \zeta_3 \right)
		  + \lamV^4 \left(\frac{897}{2} + 252 \zeta_3 \right)
		  + (\lamV \cdot \Lambda^2 \cdot \lamV) \left(459 + 396 \zeta_3 \right)
		  .\label{eq:blamS3l}
\end{align}
\label{eq:blamS}
\end{subequations}
}
{\allowdisplaybreaks 
\begin{subequations}
\begin{align}
\beta^{(1)}_{\lamV} & = 
6 \left(  \lamS \cdot \lamV + (\Lambda \cdot \lamV) \right) 
	,\label{eq:blamV1l}\\
\beta^{(2)}_{\lamV} & = 
		\lamV \left( 
			- \frac{7}{2} \trlam{}{2} 
			+ \frac{1}{4} \trlam{2}{}	
			- \frac{5}{2} \lamS \cdot \trlam{}{}
			-\frac{127}{4} \lamS^2 
			- 39 \lamV^2 
		\right) 
		\nonumber\\
		& 
		+ (\Lambda \cdot \lamV) 
		\left(
			2 \trlam{}{}
			- 51 \lamS
		\right)
		- \frac{61}{2} (\Lambda^2 \cdot \lamV)
	,\label{eq:blamV2l} \\
\beta^{(3)}_{\lamV} & = 
	\lamV \left[
		 \trlam{}{3} \left( \frac{841}{6} + 84 \zeta_3 \right)
		 - \trlam{}{} \cdot \trlam{}{2} \left( \frac{857}{8} + 60 \zeta_3 \right)
		 + \trlam{3}{} \left( \frac{743}{24} + 18 \zeta_3 \right.
	\right)
	\nonumber\\
	& 
		+ \lamS^3 \left( \frac{2005}{8} + 138 \zeta_3\right)  
		+ \lamS^2 \cdot \trlam{}{} \left( \frac{135}{4} + 18 \zeta_3 \right) 
		- \lamS \cdot \trlam{2}{} \left( \frac{53}{4} + 6 \zeta_3 \right)
	\nonumber\\
	& 
	\left.
	+ \lamS \cdot \trlam{}{2} \left( \frac{871}{8} + 60 \zeta_3 \right)
	+ \lamS \cdot \lamV^2 \left( 897 + 504 \zeta_3 \right)
	%+ 0 \cdot \trlam{}{} \cdot \lamV^2  
	+ (\lamV \cdot \Lambda \cdot \lamV) 
	\left( 444 + 252 \zeta_3 \right) 
	\right]
	\nonumber \\
	& 
	+ (\Lambda \cdot \lamV) \left[ 
	   \lamS^2 \left( \frac{4665}{8} + 324 \zeta_3 \right)	
	 - \lamS \cdot \trlam{}{}  \left( \frac{71}{4} + 24 \zeta_3 \right)
		\right.
	\nonumber \\
	& \left. 
	+ \frac{93}{2} \trlam{}{2}
	 - \trlam{2}{} \left( \frac{687}{8} + 48 \zeta_3 \right)
	+ \lamV^2 \left( 453 + 252 \zeta_3 \right)
	\right]
	\nonumber \\
	& 
	+ (\Lambda^2 \cdot \lamV) \left[ 
		\lamS \left( \frac{1255}{2} + 384 \zeta_3 \right)
	 	+ \trlam{}{} \left( \frac{393}{2} + 120 \zeta_3 \right)
	\right]
	.\label{eq:blamV3l}
\end{align}
\label{eq:blamV}
\end{subequations}
}
{\allowdisplaybreaks
\begin{subequations}
\begin{align}
\beta^{(1)}_{\lamT} & = 
	\lamT \left( 3 \lamS - \trlam{}{} \right)
	+ 4 \lamT^2 
	+ 6 (\lamV \otimes \lamV)
	,\label{eq:blamT1l}\\
\beta^{(2)}_{\lamT} & = 
		\lamT \left[
		  \frac{3}{2} \trlam{}{2}
		+ \frac{13}{4} \trlam{2}{}
		+ \frac{7}{2} \lamS \cdot \trlam{}{} 
		- \frac{61}{4} \lamS^2 
		- 7 \lamV^2 
		\right]
		\nonumber\\
	& 
		- \lamT^2 \left[ 
			24 \lamS
			+ 8 \trlam{}{}
		\right]
		+
		(\lamV \otimes \lamV) \left(
		2 \trlam{}{} - 51 \lamS
		\right)
	\nonumber\\
	&
		- \frac{61}{2}
		\left(
			\lamT \cdot \lamV \otimes \lamV 
			+ \lamV \otimes \lamT \cdot \lamV 
		\right)
	+ 
		\left( 
			  4 \trlam{}{} \cdot \trlam{}{2}
			  - \frac{4}{3} \trlam{3}{}
			  - \frac{8}{3} \trlam{}{3}
		\right)
	,\label{eq:blamT2l}
	\\
% changing to new basis:
\beta^{(3)}_{\lamT} & = 
		\lamT \left[
			37 \trlam{}{3} 
		+ \frac{29}{2} \trlam{}{} \cdot \trlam{}{2}
		- \trlam{3}{} \left( \frac{327}{16} + 6 \zeta_3 \right)
		+ \frac{201}{2} \lamS \cdot \trlam{}{2}
		\right.
		\nonumber\\
	&  \left.
		- \lamS \cdot \trlam{2}{} \left( \frac{1521}{16} + 42 \zeta_3\right) 
		- \lamS^2 \cdot \trlam{}{} 
		\left( 
			\frac{533}{16} + 18 \zeta_3 
		\right)
		+ \lamS^3 
		\left(
			\frac{1349}{16} + 66 \zeta_3 
		\right)
		\right.
		\nonumber \\
	&
		\left.
		- \trlam{}{} \cdot \lamV^2 
		\left(
			%\frac{583}{4} 
			 \frac{1291}{4}
			+ 120 \zeta_3 
		\right)
		+ \lamS \cdot \lamV^2 \left( \frac{751}{4} + 120 \zeta_3 \right)
		+ %\frac{441}{2} 
		\frac{795}{2}
		(\lamV \cdot \lamT \cdot \lamV) 
		\right]
	  \nonumber\\
	  &
	 +	
		\lamT^2 \left[ 
	  \lamS^2 \left( \frac{839}{4} + 144 \zeta_3 \right)		
	  + \lamS \cdot \trlam{}{} \left( \frac{315}{2} + 96 \zeta_3 \right) 
	  + \trlam{2}{} \left( \frac{141}{4} + 12 \zeta_3\right) 
		\right.
		\nonumber\\
	  &	\left.
	  + \lamV^2 \left( 
	  %\frac{429}{2} 
	  \frac{783}{2} 
	  + 252 \zeta_3 \right)
	  - \frac{83}{2} \trlam{}{2}
		\right]
	 +
	 (\lamV \otimes \lamV) \left[
		\lamS^2 \left( \frac{3951}{8} + 324 \zeta_3  \right)
		+ %\frac{435}{4} \trlam{}{2}
		   \frac{789}{4} \trlam{}{2}
		\right.
	\nonumber\\
	& 
	\left.
		- \lamS \cdot \trlam{}{} 
		\left( 
			\frac{143}{4} + 24 \zeta_3 
		\right)
		- \trlam{2}{} \left(
			%\frac{575}{8} 
			\frac{1283}{8} 
			+ 48 \zeta_3 
		\right)
		+ \lamV^2 \left( \frac{897}{2} + 252 \zeta_3 \right)
		\right]
	\nonumber
	\\
	&
	+ \left( \lamT \cdot \lamV \otimes \lamV + \lamV \otimes \lamT \cdot \lamV \right) 
		\left[
			\lamS \left( \frac{2333}{4} + 384 \zeta_3 \right)
			+ \trlam{}{} \left(  
			%\frac{397}{4} 
			\frac{1105}{4} 
		+ 120 \zeta_3 \right)
		\right]
	\nonumber \\
	& 
	%+ 177 
	%\left( \lamT^2 \cdot \lamV \otimes \lamV + \lamV \otimes \lamT^2 \cdot \lamV \right) 
	- 177 \left(\lamT \cdot \lamV \otimes \lamT \cdot \lamV\right)
	+ \left[
		\trlam{4}{} \left( \frac{28}{3} + 2 \zeta_3 \right)
		+ \lamS \cdot \trlam{3}{} \left( \frac{92}{3} + 16 \zeta_3 \right)
	\right.
	\nonumber\\
	& \left. 
		- \lamS \cdot \trlam{}{}\cdot \trlam{}{2}
		\left( 92 + 48 \zeta_3 \right)
		+ \left( \trlam{2}{} - \trlam{}{2} \right) \cdot \lamV^2 \left( \frac{621}{4} + 54 \zeta_3 \right) 
	  \right.
	  \nonumber \\
	  & 
	  + \trlam{}{} \cdot \trlam{}{3} \left( \frac{56}{3} + 4 \zeta_3 \right) 
		- \trlam{2}{} \cdot \trlam{}{2} \left( 28 + 6 \zeta_3 \right)
		\nonumber\\
	&
	  \left.
	  + \lamS \cdot \trlam{}{3} \left( \frac{184}{3} + 32 \zeta_3 \right)
	  + \left( \lamV \cdot \left( \lamT^2  - \trlam{}{} \cdot \lamT  
	  \right) \cdot \lamV\right) 
		  \left(
		  \frac{621}{3} + 108 \zeta_3 
	  \right)
	  \right]
	.\label{eq:blamT3l}
\end{align}
\label{eq:blamT}
\end{subequations}
}
	The mass-parameter anomalous dimensions can be cast in the following form
{\allowdisplaybreaks
\begin{subequations}
\begin{align}
\gamma_{\mS}^{(1)} & = 
	\frac{1}{2} \mS \left( 
	\trlam{}{} + 5 \lamS \right)
	+ 3 \lamV \cdot \mV, \label{eq:bmSl1} \\
\gamma_{\mS}^{(2)} & = 
	\mS \left( 
		\frac{5}{8} \trlam{2}{}
		-\frac{15}{4} \trlam{}{2}
		-\frac{25}{8} \lamS^2
		-\frac{5}{4} \lamS \cdot \trlam{}{} 
		-\frac{15}{2} \lamV \cdot \lamV
		 \right)
	\nonumber\\
	& 
		 - \frac{15}{2} 
		 \left( 
			 \lamS\, \lamV \cdot \mV 
			 + (\lamV \cdot \lamT \cdot \mV)
		 \right),
\label{eq:bmSl2} \\
\gamma_{\mS}^{(3)} & = 
		\mS \left(
			\frac{81}{2} \trlam{}{3} 
			-\frac{117}{8} \trlam{}{} \cdot \trlam{}{2}  
			+ \frac{51}{32} \cdot \trlam{3}{}
			+ \frac{1155}{32} \lamS^3 + \frac{693}{32} \lamS^2 \cdot \trlam{}{}
			\right.
			\nonumber\\
			& \left.- \frac{171}{32} \lamS \cdot \trlam{2}{}
			+ \frac{387}{8} \lamS \cdot \trlam{}{2}
			+ \frac{1557}{8} \lamS \cdot \lamV^2 
			+ \frac{9}{8} \trlam{}{} \cdot \lamV^2
			+ 189 (\lamV \cdot \lamT \cdot \lamV)
		\right) 
		\nonumber\\
		& 
	+ \frac{441}{4} (\lamV \cdot \lamT^2 \cdot \mV)
		+ \lamV \cdot \mV \left(
		   \frac{1659}{16} \lamS^2 
		 + \frac{51}{8} \lamS \trlam{}{} 
		 +  \frac{39}{16} \trlam{2}{}
		 -\frac{9}{2} \trlam{}{2}
	\right)
	\nonumber\\
	&
	+ \left( \frac{351}{2} \lamS - 9 \trlam{}{}
	\right) (\lamV \cdot \lamT \cdot \mV).
\label{eq:bmSl3}
\end{align}
\label{eq:bmS}
\end{subequations}
}
{\allowdisplaybreaks 
\begin{subequations}
\begin{align}
\gamma_{\mV}^{(1)} & = 
	\frac{1}{2} \mV \left( \lamS - \trlam{}{} \right)
	+ 3 \left( \mS \lamV + \lamT \cdot \mV\right),
\label{eq:bmV1l}
\\
\gamma_{\mV}^{(2)} & = 
	\mV \left(
		\frac{9}{4} \trlam{}{2}
		- \frac{3}{8} \trlam{2}{}
		- \frac{9}{8} \lamS^2
		+\frac{3}{4} \lamS \cdot \trlam{}{}
		+ \frac{3}{2} \lamV \cdot \lamV
	\right)
	\nonumber\\
	& 
	- \lamV \left( 
		\frac{15}{2} \lamS \mS
		+ 9 \lamV \cdot \mV
	\right)
	- \frac{15}{2} \mS \, (\lamT \cdot \lamV)
	- 6 \lamS \, (\lamT \cdot \mV)
	- 3 (\lamT^2 \cdot \mV), 
\label{eq:bmV2l}
	\\
\gamma_{\mV}^{(3)} & = 
		\mV \left[
			\frac{17}{2} \trlam{}{3} 
			-\frac{21}{4} \trlam{}{} \cdot \trlam{}{2}  
			+ \frac{57}{32} \cdot \trlam{3}{}
			+ \frac{199}{32} \lamS^3 
			- \frac{273}{32} \lamS^2 \cdot \trlam{}{}
			\right.
			\nonumber\\
			& \left.+ \frac{21}{32} \lamS \cdot \trlam{2}{}
			- \frac{63}{8} \lamS \cdot \trlam{}{2}
			+ \frac{27}{8} \lamS \cdot \lamV^2 
			- \frac{33}{8} \trlam{}{} \cdot \lamV^2
			- \frac{51}{4} (\lamV \cdot \lamT \cdot \lamV)
		\right] \nonumber\\
	& + \mS\, \lamV 
		 \left[\frac{27}{2}  \trlam{}{2}
		- \frac{9}{16}  \trlam{2}{}
		+ \frac{81}{8}  \lamS \trlam{}{}
		+ \frac{1683}{16}  \lamS^2
		+ \frac{513}{4}  \lamV^2 
			\right]
			\nonumber\\
	& + \lamV
			\left[
(\lamV\cdot\mV)
		\left(
				3 \trlam{}{}  
		+ \frac{747}{4} \lamS 
			\right) 
		+ \frac{303}{4} (\lamV \cdot \lamT \cdot \mV)
		\right]
	 + 108 \mS \, (\lamT^2 \cdot \lamS)	
		\nonumber\\
	& 
		+ (\lamT \cdot \mV)
		\left[
			  \frac{981}{16} \lamS^2
			-\frac{153}{8} \trlam{}{2}
			-\frac{57}{16} \trlam{2}{}
			-\frac{63}{8} \lamS \trlam{}{}
			+ \frac{147}{4} \lamV^2
		\right]
	 + 21 \,\trlam{}{} \, (\lamT^2 \cdot \mV)
		\nonumber \\
	& 
		+ (\lamT \cdot \lamV)
		\left[
			\frac{639}{4} \lamS \, \mS
			- \frac{45}{4} \trlam{}{} \, \mS
			+ 96 (\lamV \cdot \mV)
		\right]
	 + 81 \lamS \, (\lamT^2 \cdot \mV).
\label{eq:bmVl3}
\end{align}
\label{eq:bmV}
\end{subequations}
}
By means of simple algebra one can easily convert these results to the expressions for RG functions $\beta_{\lambda_i}$, $\gamma_{m^2_{ij}}$ of the initial Higgs potential \eqref{eq:V_gen} (see next section) or to the beta-functions for reparametrization invariants (see appendix~\ref{app:invariants_rge}) .

\section{\label{sec:rge2_lambda_m}RGE for $\lambda_i$ and $m^2_{ij}$}

We define the RG functions of the parameters $\lambda_i$ and $m^2_{ij}$ from the potential given in eq.~\eqref{eq:V_gen} as: 
\begin{align}
	\frac{d \lambda_i}{d \ln \mu^2} & = 
	\frac{1}{2} \frac{d \lambda_i}{d \ln \mu}  = 
	\sum_{l=1}^\infty h^l \beta^{(l)}_{\lambda_i}, 
	\quad
	\frac{d m^2_{ij}}{d \ln \mu^2}  = 
	\frac{1}{2} \frac{d m^2_{ij}}{d \ln \mu}  = 
	\sum_{l=1}^\infty h^l \gamma^{(l)}_{m^2_{ij}}.
\label{eq:beta_l_m_def}
\end{align}
Having in mind that
	\begin{align}
		\lambda_1 & = \frac{\lamS + \lamT_{33}}{2} + (\lamV)_3,
		\quad
		\lambda_2  = \frac{\lamS + \lamT_{33}}{2} - (\lamV)_3,
		\quad
		\lambda_3  = \frac{\lamS - \lamT_{33}}{2}, \\
		\lambda_4 & = \frac{\lamT_{11} + \lamT_{22}}{2},
		\qquad	
		\lambda_5  = \frac{\lamT_{11} - \lamT_{22}}{2} - i \lamT_{12},
		\\
		\lambda_6 & = \frac{(\lamV)_1+\lamT_{13}}{2} -i \frac{(\lamV)_2 + \lamT_{23}}{2},
		\qquad
		\lambda_7  = \frac{(\lamV)_1-\lamT_{13}}{2} -i \frac{(\lamV)_2 - \lamT_{23}}{2},
	\end{align}
	one can obtain the three-loop results for $\beta_{\lambda_i}$ \eqref{eq:beta_l_m_def}.
For brevity we present here only one- and two-loop contributions\footnote{Full three-loop result is available in computer-readable form as an ancillary file of the arXiv version of the paper.}:
\begin{align}
	2\beta^{(1)}_{\lambda_1} & = 
    12\lambda_1{}^2
    +2\left[\lambda_4{}^2+\left|\lambda_5\right|{}^2\right]
    +4\left[\lambda_3 \lambda_4+\lambda_3{}^2\right]
    +24\left|\lambda_6\right|{}^2,
    \label{eq:bet1_L1}
    \\
	2\beta^{(2)}_{\lambda_1} & = 
    -78\lambda_1{}^3
    -312\lambda_1 \left|\lambda_6\right|{}^2
    -134\lambda_4 \left|\lambda_6\right|{}^2
    -126\lambda_3 \left|\lambda_6\right|{}^2
    %-71\left[\lambda_5 \left(\lambda_{6}^{*}\right){}^2+\lambda_6{}^2 \left(\lambda_{5}^{*}\right)\right]
    -142 \,\Re \left[\lambda_5 \left(\lambda_{6}^{*}\right){}^2\right] %+\lambda_6{}^2 \left(\lambda_{5}^{*}\right)\right]
    -44\lambda_4 \left|\lambda_5\right|{}^2
   \nonumber\\
   &
    -40\lambda_3 \left|\lambda_5\right|{}^2
    -36\lambda_3 \left|\lambda_7\right|{}^2
    %-33\left[\lambda_3 \lambda_6 \left(\lambda_{7}^{*}\right)+\lambda_3 \lambda_7 \left(\lambda_{6}^{*}\right)\right]
    -66 \,\Re \left[\lambda_3 \lambda_6 \left(\lambda_{7}^{*}\right)\right] %+\lambda_3 \lambda_7 \left(\lambda_{6}^{*}\right)\right]
    -32\lambda_3 \lambda_4{}^2
    -28\lambda_4 \left|\lambda_7\right|{}^2
    %-25\left[\lambda_4 \lambda_6 \left(\lambda_{7}^{*}\right)+\lambda_4 \lambda_7 \left(\lambda_{6}^{*}\right)\right]
    -50 \, \Re \left[\lambda_4 \lambda_6 \left(\lambda_{7}^{*}\right)\right] %+\lambda_4 \lambda_7 \left(\lambda_{6}^{*}\right)\right]
   \nonumber	\\
   &
    -24\lambda_3{}^2 \lambda_4
    %-20\left[\lambda_1 \lambda_3 \lambda_4+\lambda_1 \lambda_3{}^2\right]
    -20\lambda_1 \lambda_3 \left[ \lambda_4+ \lambda_3\right]
    %-17\left[\lambda_5 \left(\lambda_{6}^{*}\right) \left(\lambda_{7}^{*}\right)+\lambda_6 \lambda_7 \left(\lambda_{5}^{*}\right)\right]
    -34 \, \Re \left[\lambda_6 \lambda_7 \left(\lambda_{5}^{*}\right)\right]
    -16\lambda_3{}^3
    -14\lambda_1 \left|\lambda_5\right|{}^2
   \nonumber	\\
   &
    %-12\left[\lambda_1 \lambda_4{}^2+\lambda_4{}^3\right]
    -12 \lambda_4^2 \left[\lambda_1 +\lambda_4\right]
    %-10\left[\lambda_5 \left(\lambda_{7}^{*}\right){}^2+\lambda_7{}^2 \left(\lambda_{5}^{*}\right)\right]
    -20 \, \Re \left[\lambda_5 \left(\lambda_{7}^{*}\right){}^2\right] % +\lambda_7{}^2 \left(\lambda_{5}^{*}\right)\right]
    %+3\left[\lambda_2 \lambda_6 \left(\lambda_{7}^{*}\right)+\lambda_2 \lambda_7 \left(\lambda_{6}^{*}\right)\right]
    +6 \lambda_2\,\Re \left[\lambda_6 \left(\lambda_{7}^{*}\right)\right] %+\lambda_2 \lambda_7 \left(\lambda_{6}^{*}\right)\right]
    +6\lambda_1 \left|\lambda_7\right|{}^2,
    \label{eq:bet2_L1}
\end{align}
\begin{align}
	2\beta^{(1)}_{\lambda_3} & = 
	2\left[\lambda_4 (\lambda_1 + \lambda_2)+\lambda_4{}^2+\left|\lambda_5\right|{}^2\right]
    +4\left[\lambda_3{}^2+\left|\lambda_6\right|{}^2+\left|\lambda_7\right|{}^2\right]
    \nonumber\\
    &
    +6 \lambda_3 \left[\lambda_1 +\lambda_2\right]
    %+8\left[\lambda_6 \left(\lambda_{7}^{*}\right)+\lambda_7 \left(\lambda_{6}^{*}\right)\right]
    +16 \, \Re\left[\lambda_6 \left(\lambda_{7}^{*}\right) \right], %+\lambda_7 \left(\lambda_{6}^{*}\right)\right]
    \label{eq:bet1_L3}
    \\
	2\beta^{(1)}_{\lambda_3} & = 
    %-85\left[\lambda_3 \lambda_6 \left(\lambda_{7}^{*}\right)+\lambda_3 \lambda_7 \left(\lambda_{6}^{*}\right)\right]
-170 \lambda_3 \, \Re \left[ \lambda_6 \left(\lambda_{7}^{*}\right)\right]
    -65 \lambda_4 \left[\left|\lambda_6\right|{}^2+ \left|\lambda_7\right|{}^2\right]
    -59\left[\lambda_1 \left|\lambda_6\right|{}^2+\lambda_2 \left|\lambda_7\right|{}^2\right]
   \nonumber\\
   &
    -44\lambda_4 \left|\lambda_5\right|{}^2
    %-41\left[\lambda_4 \lambda_6 \left(\lambda_{7}^{*}\right)+\lambda_4 \lambda_7 \left(\lambda_{6}^{*}\right)\right]
    -82 \lambda_4 \Re \, \left[\lambda_6 \left(\lambda_{7}^{*}\right)\right]
    -36\left[\lambda_1 \lambda_3{}^2+\lambda_2 \lambda_3{}^2\right]
    %-33\left[\lambda_5 \left(\lambda_{6}^{*}\right) \left(\lambda_{7}^{*}\right)+\lambda_6 \lambda_7 \left(\lambda_{5}^{*}\right)\right]
    -66 \Re\, \left[\lambda_6 \lambda_7 \left(\lambda_{5}^{*}\right)\right]
   \nonumber\\
   &
    %-\frac{65}{2}\left[\lambda_5 \left(\lambda_{6}^{*}\right){}^2+\lambda_5 \left(\lambda_{7}^{*}\right){}^2+\lambda_6{}^2 \left(\lambda_{5}^{*}\right)+\lambda_7{}^2 \left(\lambda_{5}^{*}\right)\right]
    -65 \, \Re \left[\lambda_5 \left(\lambda_{6}^{*}\right){}^2+\lambda_5 \left(\lambda_{7}^{*}\right){}^2\right] %+\lambda_6{}^2 \left(\lambda_{5}^{*}\right)+\lambda_7{}^2 \left(\lambda_{5}^{*}\right)\right]
    -22\left[\lambda_1 \left|\lambda_7\right|{}^2+\lambda_2 \left|\lambda_6\right|{}^2\right]
    %-\frac{41}{2}\left[\lambda_1 \lambda_6 \left(\lambda_{7}^{*}\right)+\lambda_1 \lambda_7 \left(\lambda_{6}^{*}\right)+\lambda_2 \lambda_6 \left(\lambda_{7}^{*}\right)+\lambda_2 \lambda_7 \left(\lambda_{6}^{*}\right)\right]
    - 41 (\lambda_1 + \lambda_2) \Re \left[ \lambda_6 \left(\lambda_{7}^{*}\right)\right] %+\lambda_1 \lambda_7 \left(\lambda_{6}^{*}\right)+\lambda_2 \lambda_6 \left(\lambda_{7}^{*}\right)+\lambda_2 \lambda_7 \left(\lambda_{6}^{*}\right)\right]
   \nonumber\\
   &
    -18\left|\lambda_5\right|{}^2\left[\lambda_1 +\lambda_2 +\lambda_3 \right]
    -16 \lambda_3 \lambda_4 \left[\lambda_1 +\lambda_2 +\lambda_4\right]
    -15 \lambda_3 \left[\lambda_1{}^2 +\lambda_2{}^2\right]
    -14 \lambda_4^2 \left[\lambda_1+\lambda_2\right]
   \nonumber\\
   &
    -12\left[\lambda_3{}^3+\lambda_4{}^3\right]
    -4 \lambda_4 \left[\lambda_1{}^2 +\lambda_2{}^2 +\lambda_3{}^2 \right]
    -57 \lambda_3 \left[ \left|\lambda_6\right|{}^2+ \left|\lambda_7\right|{}^2\right],
    \label{eq:bet2_L3}
\end{align}
\begin{align}
	2\beta^{(1)}_{\lambda_4} & = 
    2\left[\lambda_4 (\lambda_1+\lambda_2)
    %+\lambda_6 \left(\lambda_{7}^{*}\right)+\lambda_7 \left(\lambda_{6}^{*}\right)
	    + 2 \Re \,\left[ \lambda_6 \left(\lambda_{7}^{*}\right)\right]
    \right]
    +4\lambda_4{}^2
    \nonumber\\
    &
    +8\left[\lambda_3 \lambda_4+\left|\lambda_5\right|{}^2\right]
    +10\left[\left|\lambda_6\right|{}^2+\left|\lambda_7\right|{}^2\right],
    \label{eq:bet1_L4}
    \\
	2\beta^{(2)}_{\lambda_4} & = 
    %-77\lambda_4 (\lambda_6 \left(\lambda_{7}^{*}\right)+\lambda_7 \left(\lambda_{6}^{*}\right))
	-154 \lambda_4 \Re \,[\lambda_6 \left(\lambda_{7}^{*}\right)]
    -71\left[\lambda_1 \left|\lambda_6\right|{}^2+\lambda_2 \left|\lambda_7\right|{}^2\right]
    -69\lambda_3 \left(\left|\lambda_6\right|{}^2+\left|\lambda_7\right|{}^2\right)
    \nonumber \\
    &
    -65\lambda_4 \left(\left|\lambda_6\right|{}^2+\left|\lambda_7\right|{}^2\right)
    -48\lambda_3 \left|\lambda_5\right|{}^2
    %-45\left[\lambda_5 \left(\lambda_{6}^{*}\right) \left(\lambda_{7}^{*}\right)+\lambda_6 \lambda_7 \left(\lambda_{5}^{*}\right)\right]
    -90 \Re \, \left[\lambda_6 \lambda_7 \left(\lambda_{5}^{*}\right)\right]
    -40\lambda_3 \lambda_4 (\lambda_1+\lambda_2)
    \nonumber \\
    &
    %-\frac{77}{2}\left[\lambda_5 \left(\left(\lambda_{6}^{*}\right){}^2+\left(\lambda_{7}^{*}\right){}^2\right)+\left(\lambda_{5}^{*}\right) \left(\lambda_6{}^2+\lambda_7{}^2\right)\right]
    -77\Re\,\left[\left(\lambda_{5}^{*}\right) \left(\lambda_6{}^2+\lambda_7{}^2\right)\right]
    %-37\lambda_3 (\lambda_6 \left(\lambda_{7}^{*}\right)+\lambda_7 \left(\lambda_{6}^{*}\right))
    -74 \lambda_3 \Re\, \left[\lambda_6 \left(\lambda_{7}^{*}\right)\right]
    -28\lambda_3 \lambda_4 (\lambda_3+\lambda_4)
    \nonumber \\
    &
    -26\lambda_4 \left|\lambda_5\right|{}^2
    -24\left|\lambda_5\right|{}^2 (\lambda_1+\lambda_2)
    -20\lambda_4{}^2 (\lambda_1+\lambda_2)
    -10\left[\lambda_1 \left|\lambda_7\right|{}^2+\lambda_2 \left|\lambda_6\right|{}^2\right]
    \nonumber \\
    &
    %-\frac{17}{2}(\lambda_1+\lambda_2) (\lambda_6 \left(\lambda_{7}^{*}\right)+\lambda_7 \left(\lambda_{6}^{*}\right))
    -17(\lambda_1+\lambda_2) \Re \, \left[\lambda_6 \left(\lambda_{7}^{*}\right)\right]
    -7\lambda_4 \left(\lambda_1{}^2+\lambda_2{}^2\right),
    \label{eq:bet2_L4}
\end{align}
\begin{align}
	2\beta^{(1)}_{\lambda_5} & = 
    2\lambda_5 (\lambda_1+\lambda_2)
    +4\lambda_6 \lambda_7
    +8\lambda_3 \lambda_5
    +10\left[\lambda_6{}^2+\lambda_7{}^2\right]
    +12\lambda_4 \lambda_5,
    \label{eq:bet1_L5}
    \\
	2\beta^{(2)}_{\lambda_5} & = 
    -82\lambda_4 \lambda_6 \lambda_7
    %-81\lambda_5 (\lambda_6 \left(\lambda_{7}^{*}\right)+\lambda_7 \left(\lambda_{6}^{*}\right))
    -192\lambda_5 \Re\, \left[\lambda_6 \left(\lambda_{7}^{*}\right)\right]    -76\lambda_3 \lambda_4 \lambda_5
    -74\lambda_3 \lambda_6 \lambda_7
    -73\lambda_4 \left(\lambda_6{}^2+\lambda_7{}^2\right)
    \nonumber \\
    &
    -71\left[\lambda_1 \lambda_6{}^2+\lambda_2 \lambda_7{}^2\right]
    -69\left[\lambda_3 \left(\lambda_6{}^2+\lambda_7{}^2\right)+\lambda_5 \left|\lambda_6\right|{}^2+\lambda_5 \left|\lambda_7\right|{}^2\right]
    -44\lambda_4 \lambda_5 (\lambda_1+\lambda_2)
    \nonumber \\
    &
    -40\lambda_3 \lambda_5 (\lambda_1+\lambda_2)
    -32\lambda_4{}^2 \lambda_5
    -28\lambda_3{}^2 \lambda_5
    -17\lambda_6 \lambda_7 (\lambda_1+\lambda_2)
    \nonumber \\
    &
    -10\left[\lambda_1 \lambda_7{}^2+\lambda_2 \lambda_6{}^2\right]
    -7\lambda_5 \left(\lambda_1{}^2+\lambda_2{}^2\right)
    +6\lambda_5 \left|\lambda_5\right|{}^2,
    \label{eq:bet2_L5}
\end{align}
\begin{align}
	2\beta^{(1)}_{\lambda_6} & = 
    2\lambda_5 \left(\lambda_{7}^{*}\right)
    +4\lambda_4 \lambda_7
    +6\lambda_3 (\lambda_6+\lambda_7)
    +8\lambda_4 \lambda_6
    +10\lambda_5 \left(\lambda_{6}^{*}\right)
    +12\lambda_1 \lambda_6,
    \label{eq:bet1_L6}
    \\
	2\beta^{(2)}_{\lambda_6} & = 
    -111\lambda_6 \left|\lambda_6\right|{}^2
    -84\lambda_7 \left|\lambda_6\right|{}^2
    -78\lambda_1{}^2 \lambda_6
    -73\lambda_4 \lambda_5 \left(\lambda_{6}^{*}\right)
    -71\lambda_1 \lambda_5 \left(\lambda_{6}^{*}\right)
    \nonumber \\
    &
    -69\lambda_3 \lambda_5 \left(\lambda_{6}^{*}\right)
    -67\lambda_1 \lambda_4 \lambda_6
    -65\lambda_3 \lambda_4 \lambda_6
    -63\lambda_1 \lambda_3 \lambda_6
    -53\lambda_3 \lambda_4 \lambda_7
    \nonumber \\
    &
    -42\left[\lambda_6{}^2 \left(\lambda_{7}^{*}\right)+\lambda_7 \left|\lambda_7\right|{}^2\right]
    -41\lambda_4 \lambda_5 \left(\lambda_{7}^{*}\right)
    -\frac{81}{2}\lambda_7 \left|\lambda_5\right|{}^2
    -37\lambda_3 \lambda_5 \left(\lambda_{7}^{*}\right)
    \nonumber \\
    &
    -\frac{69}{2}\left[\lambda_3{}^2 \lambda_7+\lambda_6 \left|\lambda_5\right|{}^2\right]
    -\frac{65}{2}\lambda_4{}^2 (\lambda_6+\lambda_7)
    -\frac{61}{2}\lambda_3{}^2 \lambda_6
    -22\lambda_7{}^2 \left(\lambda_{6}^{*}\right)
    \nonumber \\
    &
    -18\lambda_2 \lambda_3 \lambda_6
    -\frac{33}{2}\lambda_3 \lambda_7 (\lambda_1+\lambda_2)
    -14\lambda_2 \lambda_4 \lambda_6
    -\frac{25}{2}\lambda_4 \lambda_7 (\lambda_1+\lambda_2)
    -11\lambda_6 \left|\lambda_7\right|{}^2
    \nonumber \\
    &
    -10\lambda_2 \lambda_5 \left(\lambda_{6}^{*}\right)
    -\frac{17}{2}\lambda_5 \left(\lambda_{7}^{*}\right) (\lambda_1+\lambda_2)
    +\frac{3}{2}\lambda_2 (\lambda_1 \lambda_7+\lambda_2 \lambda_6).
    \label{eq:bet2_L6}
\end{align}

The expressions for $\beta_{\lambda_{2}}$ ($\beta_{\lambda_7}$) can be obtained from that of $\beta_{\lambda_{1}}$ ($\beta_{\lambda_6}$) via the substitutions $\lambda_1 \leftrightarrow \lambda_2$ and $\lambda_6 \leftrightarrow \lambda_7$.   

	The anomalous dimensions of the mass parameters \eqref{eq:V_gen} 
\begin{align}
	m_{11}^2 & = \frac{1}{2}
	\left[\mS + (\mV)_3\right], 
\quad 
m_{22}^2  = \frac{1}{2} \left[ \mS - (\mV)_3\right],
	\quad 
	m_{12}^2  = \frac{1}{2}\left[ - (\mV)_1 + i (\mV)_2 \right] 
\end{align}
	can be cast into 
{\allowdisplaybreaks
\begin{align}
	2 \gamma^{(1)}_{m_{11}^2} & = 
	6\lambda_1\,m_{11}^2
	+ \left(2\lambda_4+4\lambda_3\right)\,m_{22}^2 -6\lambda_6\,m_{12}^{*2} -6\left(\lambda_{6}^{*}\right)\,m_{12}^2,
	\label{eq:gam1_mm11}
	\\
	2 \gamma^{(2)}_{m_{11}^2} & = 
	\left(3\left|\lambda_7\right|{}^2-27\left|\lambda_6\right|{}^2-15\lambda_1{}^2-3\left|\lambda_5\right|{}^2-2\left[\lambda_3 \lambda_4+\lambda_3{}^2+\lambda_4{}^2\right]\right)\,m_{11}^2 \nonumber\\
				  &
    - \left(18\left[\left|\lambda_6\right|{}^2+\left|\lambda_7\right|{}^2\right]+12\left|\lambda_5\right|{}^2+8\left[\lambda_3 \lambda_4+\lambda_3{}^2+\lambda_4{}^2\right]\right)\,m_{22}^2 \nonumber\\
    &
    + \left[ \left(\frac{9}{2}\left[\lambda_7 (\lambda_3 +\lambda_4) +\lambda_5 \left(\lambda_{7}^{*}\right)\right]
+\frac{21}{2}\left[\lambda_6 (\lambda_3+\lambda_4 )+\lambda_5 \left(\lambda_{6}^{*}\right)\right]
    \right. \right.
    \nonumber\\
    & \left. \left. -\frac{3}{2}\lambda_2 \lambda_7+\frac{33}{2}\lambda_1 \lambda_6\right)\,m_{12}^{*2} +\! \mathrm{h.c.}\right]\!\!,%\\
    \label{eq:gam2_mm11}
\\
%	2 \gamma^{(1)}_{m_{22}^2} & = 
%    6\lambda_2\,m_{22}^2
%    + \left(2\lambda_4+4\lambda_3\right)\,m_{11}^2
%    -6\lambda_7\,m_{12}^{*2}
%    -6\left(\lambda_{7}^{*}\right)\,m_{12}^2,
%    \label{eq:gam1_mm22}
%    \\
%	2 \gamma^{(2)}_{m_{22}^2} & = 
%     \left(3\left|\lambda_6\right|{}^2-27\left|\lambda_7\right|{}^2-15\lambda_2{}^2-3\left|\lambda_5\right|{}^2-2\left[\lambda_3 \lambda_4+\lambda_3{}^2+\lambda_4{}^2\right]\right)\,m_{22}^2 \nonumber\\
%    & - \left(18\left[\left|\lambda_6\right|{}^2+\left|\lambda_7\right|{}^2\right]+12\left|\lambda_5\right|{}^2+8\left[\lambda_3 \lambda_4+\lambda_3{}^2+\lambda_4{}^2\right]\right)\,m_{11}^2\nonumber\\
%    &+
%     \left[\left(\frac{9}{2}\left[\lambda_6 (\lambda_3 +\lambda_4 )+\lambda_5 \left(\lambda_{6}^{*}\right)\right]-\frac{3}{2}\lambda_1 \lambda_6+\frac{21}{2}\left[\lambda_7 (\lambda_3+\lambda_4)+\lambda_5 \left(\lambda_{7}^{*}\right)\right]+\frac{33}{2}\lambda_2 \lambda_7\right)\,m_{12}^{*2} + \mathrm{h.c.}\right],
%\label{eq:gam2_mm22}
%\\
	2 \gamma^{(1)}_{m_{12}^2} & = 
    \left(2\lambda_3+4\lambda_4\right)\,m_{12}^2
    -6 \left(\lambda_6\,m_{11}^2 +\lambda_7\,m_{22}^2 -\lambda_5\,m_{12}^{*2} \right),
    \label{eq:gam1_mm12}
    \\
	2 \gamma^{(2)}_{m_{12}^2} & = 
	\left(
	\frac{3}{2}\left[\lambda_1{}^2+\lambda_2{}^2\right]	
		-12\left[\lambda_6 \left(\lambda_{7}^{*}\right)+\lambda_7 \left(\lambda_{6}^{*}\right)\right]-6\left[(\lambda_1+\lambda_2) (\lambda_3+\lambda_4)+\lambda_3 \lambda_4\right]
     \right.
     \nonumber\\
     & \left. 
		\vphantom{\frac{3}{2}}
     + 3\left|\lambda_5\right|{}^2 
     \right)\,m_{12}^2 %\nonumber\\ & 
     -\left(12\left[\lambda_3 \lambda_5+\lambda_4 \lambda_5+\lambda_6 \lambda_7+\lambda_6{}^2+\lambda_7{}^2\right]+6\left[\lambda_1 \lambda_5+\lambda_2 \lambda_5\right]\right)\,m_{12}^{*2}\nonumber\\
    &
    + \left(\frac{9}{2}\left[\lambda_6(\lambda_3 +\lambda_4) +\lambda_5 \left(\lambda_{6}^{*}\right)\right]+\frac{21}{2}\left[\lambda_7 (\lambda_3 + \lambda_4) +\lambda_5 \left(\lambda_{7}^{*}\right)\right]
    \right. \nonumber \\
    & \left.
	-\frac{3}{2}\lambda_1 \lambda_6
    +\frac{33}{2}\lambda_2 \lambda_7\right)\,m_{22}^2 %\nonumber\\ &
    + \left(\frac{9}{2}\left[\lambda_7 (\lambda_3 + \lambda_4)+\lambda_5 \left(\lambda_{7}^{*}\right)\right]
-\frac{3}{2}\lambda_2 \lambda_7
    \right. \nonumber \\
    & \left.
    +\frac{21}{2}\left[\lambda_6(\lambda_3 + \lambda_4)+\lambda_5 \left(\lambda_{6}^{*}\right)\right]+\frac{33}{2}\lambda_1 \lambda_6\right)\,m_{11}^2.
\end{align}
}
Again, $\gamma_{m^2_{22}}$ can be deduced from $\gamma_{m^2_{11}}$ if we substitute $\lambda_1 \leftrightarrow \lambda_2$, $\lambda_6 \leftrightarrow \lambda_7$ and $m_{11}^2 \leftrightarrow m_{22}^2$. It is worth noting that the one-loop results presented here coincide with that given in ref.~\cite{Haber:1993an,Branco:2011iw}. In the case of real $\lambda_{1-5}$, $m_{12}^2$ and vanishing $\lambda_6=\lambda_7$ our two-loop expressions coincide with those presented in ref.~\cite{Chowdhury:2015yja} (2HDM with soft $Z_2$ breaking). 

Before going to conclusions, %of the RGEs for reparametrization invariants, %  three-loop expressions for the RG functions, which were obtained from direct diagram computations, 
let us briefly comment on the peculiar fact about well-known two-loop result \cite{Machacek:1984zw,Luo:2002ti} for beta-function of quartic coupling $\lambda_{abcd}$ (see eq.~(4.2) of ref.~\cite{Machacek:1984zw} or eq. (37) from ref.~\cite{Luo:2002ti}) in a general renormalizable QFT model:
\begin{align}
	\beta_{abcd} = \gamma_{abcd} + \sum_i \gamma^S(i) \lambda_{abcd}.
	\label{eq:beta_abcd}
\end{align}
Here, $\gamma_{abcd}$ is the anomalous dimension of the operator $\phi_a \phi_b \phi_c \phi_d$, while $\gamma^S_i$ is said to be ``the anomalous dimension of the scalar field $i$''. The subtlety we encountered is the interpretation of the last term in eq.~\eqref{eq:beta_abcd} that comes from the renormalization of the fields $\phi_a$, which suppose to carry a (gauge) index $a$. In general, the anomalous-dimension matrix of scalar fields is non-diagonal $\gamma^S_{ab}$ and the mixing due to dimension-4 operators has to be taken into account. For example, for  non-zero $\lambda_6$, $\lambda_7$ the two-loop propagator corrections in 2HDM give rise to the mixing between $\Phi_1$ and $\Phi_2$ (see also ref.~\cite{Ginzburg:2008kr}) . It is interesting to note the in the Yukawa beta-functions  possible mixing is taken into account in the general formula (see, e.g,.  eq. (32) of ref.\cite{Luo:2002ti}). However, in ref.~\cite{Luo:2002ti}  $\gamma^S_{i}$ seems to be interpreted as eigenvalues\footnote{And in public computer codes \SARAH~4.13 \cite{Staub:2013tta} and \PyRATE~2 \cite{Lyonnet:2016xiz} the sum over $i$ in \eqref{eq:beta_abcd} is replaced by the sum over diagonal elements corresponding to external legs, i.e., $\sum_{i=\{a,b,c,d\}} \gamma_{ii} $.} of $\gamma^S_{ab}$, which in our opinion leads to an incorrect result, when scalar indices $a,b,c,d$ are not related to a gauge group\footnote{We usually expect the propagators are diagonal w.r.t gauge indices.}, i.e., the expression \eqref{eq:beta_abcd} should be rewritten as 

\begin{align}
	\beta_{abcd} = \gamma_{abcd} + \left(\gamma^S_{aa'}\lambda_{a'bcd} + \mathrm{permutations}\right).
	\label{eq:beta_abcd_fixed}
\end{align}
Obviously, if $\gamma_{ab}$ is diagonal, eq.~\eqref{eq:beta_abcd_fixed} leads to eq.~\eqref{eq:beta_abcd}.
The same problem can appear in calculation of mass-parameter RGE (c.f, eq.(90) of ref.~\cite{Luo:2002ti}). 

One explicit argument for eq.~\eqref{eq:beta_abcd_fixed} comes from the
following fact. We tried to use eq.~\eqref{eq:beta_abcd} to compute two-loop $\beta_{\lambda_i}$. % for $\lambda_i$, which enter $V_H$ \eqref{eq:V_gen}. 
From the definition of $\LambdaMuNu$~\eqref{eq:LambdaMuNudef} one can easily find $\beta_{\LambdaMuNu}$, but again written in terms of $\lambda_i$. It turns out that we were not able to convert the obtained result for $\beta_{\LambdaMuNu}$ to the form (see eqs.~\eqref{eq:beta_LV_gen} and \eqref{eq:betb_LT_gen}), which involve only invariants and certain vectors/tensors constructed from $\LambdaMuNu$. On the contrary, the expression \eqref{eq:beta_abcd_fixed} gives rise to the same results \eqref{eq:blamS}, \eqref{eq:blamV} and \eqref{eq:blamT} that we derived via explicit computation of Feynman graphs. 

\section{\label{sec:conclusios}Conclusions}

We considered the three-loop RGE for the scalar sector of general 2HDM in the limit of vanishing gauge and Yukawa couplings. In spite of the fact that the obtained result is obviously incomplete and can not be used in phenomenological analysis of the model, it can be treated as a necessary step  towards full three-loop beta-functions. 

A convenient parameterization of the scalar sector was utilized to deal with combinatorics of (tensor) self-couplings and to restrict the general form of the beta-functions in the considered limit. Scalar coefficients turn out to be polynomials in a finite set of invariants. The latter give rise to a basis-independent parameterization of the Higgs sector. 

The approach can be easily extended to the case of gauge interactions since the corresponding couplings are bilinear in Higgs fields. However, the extension to the case of Yukawa interactions is not straightforward. Due to this, we will study these peculiarities elsewhere.   
It is also worth noting that one can make use of the public code \texttt{FMFT} \cite{Pikelner:2017tgv} to generalize the obtained result to the four-loop case.

An important by-product of the paper is that one should be careful, when interpreting the two-loop beta-functions of scalar self-couplings presented in refs.~\cite{Machacek:1984zw,Luo:2002ti}. We discovered that public software \cite{Staub:2013tta,Lyonnet:2016xiz}, which can be used to generate RGEs for any renormalizable model, do produce incorrect results when scalar sector with multiple higgs doublets is considered and mixing between the scalar states by dimension-4 operators is allowed. We expect that physical analyses based on RGE obtained by means of these codes may be inaccurate. 

{\bf Note added:}  Slightly after the results of the present work were made public, a paper with comprehensive study \cite{Schienbein:2018fsw} of the two-loop RGEs in general renormalizable QFT appeared on the arXiv. The authors of the reference %\cite{Schienbein:2018fsw} 
confirmed our findings and extended the analysis to the case of cubic scalar couplings and scalar mass terms\footnote{Discrepancies with previous results on fermion masses were also found in ref.~\cite{Schienbein:2018fsw}.}.  
The correct expressions involving non-diagonal anomalous dimensions of the scalar fields will be incorporated in forthcoming versions of \SARAH and \PyRATE~in the near future.

\section{Acknowledgements}
	The author would like to thank Veronika Rutberg, Andrey Pikelner, Andrey Onischenko, Mikhail Kalmykov, and Dmitri Kazakov for fruitful discussions. The work is supported in parts by the RFBR grant No. 17-02-00872-a and the Heisenberg-Landau Programme. 
\appendix
\section{\label{app:Hilbert_series}Hilbert Series and the number of Reparametrization Invariants}

A convenient way to enumerate quantities invariant under some group is the so-called Hilbert Series (see, e.g., refs.~\cite{Lehman:2015via,Jenkins:2009dy,Hanany:2010vu,Hanany:2008sb} for various applications in BSM Physics, Flavour Physics and Supersymmetric gauge theories). The series are defined as
\begin{align}
	H(t) = \sum\limits_{n=0}^\infty c_n t^n,
\end{align}
where $c_n$ gives the number of invariants of degree $n$, and $c_0=1$. 
The expression of $H(t)$ can be constructed from pure group-theoretical considerations. We develop a simple MATHEMATICA code based on the {\texttt LieART} package \cite{Feger:2012bs}  to derive the series for the invariants that can be constructed by contracting different representations of $\mathrm{SU}(2)$ group. %given the representations 
In the case of invariants, built from quartic couplings only, we obtain %$H(t)$ looks like
\begin{align}
	H(t) & = \frac{1 + t^6}{(1-t)^2 (1-t^2)^2 (1-t^3)^2 (1-t^4)}.
	\label{eq:Hilbert_series}
\end{align}	
The number of factors in the denominator ($p=7$) gives us the number of independent parameters corresponding to quartic interactions.  They are encoded in the independent invariants. The order of the invariants corresponds to the power $\alpha$ in a denominator factor $(1-t^\alpha)$. There are two of them of the order one \eqref{eq:qinv_1}%$\lamS$ and $\trlam{}{}$
, two of the order two \eqref{eq:qinv_2}%- $\trlam{}{2}$ and $\lamV \cdot \lamV$, 
, two of the order three \eqref{eq:qinv_3}%- $\trlam{}{3}$ and $\lamV \cdot \lamT \cdot \lamV$
, and one of the order four \eqref{eq:qinv_4}. %- $\lamV \cdot \lamT^2 \cdot \lamV$. 
The products of these independent invariants give rise to higher-degree invariants, the number of which can be obtained via expansion of the denominator in $t$.  The presence of numerator different from one tells us about an additional degree-six invariant, which, however, can be eliminated when raised to the second power. % \eqref{eq:I61_squared}.

One can also consider multi-graded Hilbert series for the self-couplings and introduce a separate variable $t_i$ for each irreducible representation used to construct an invariant quantity\footnote{
The expression \eqref{eq:Hilbert_series} is recovered from \eqref{eq:hilbert_series_sc} in the limit $t_i \to t$.}: %(transforming as singlets $\lamS$, $\trlam{}{}$, a triplet $\lamV$ and a five-plet $\lamT - 1/3 \cdot \trlam{}{}$)
\begin{align}
	H(\underbrace{t_{1}}_{\lamS},\underbrace{t_{2}}_{\trlam{}{}},\underbrace{t_{3}}_{\lamV},\underbrace{t_{4}}_{\tilde \Lambda})& =\frac{1+t_{3}^{3}t_{4}^{3}}{(1-t_{1})(1-t_{2})(1-t_{3}^{2})(1-t_{4}^{2})(1-t_{4}^{3})(1-t_{4}t_{3}^{2})(1-t_{3}^{2}t_{4}^{2})}.
\label{eq:hilbert_series_sc}
\end{align}
This time not only the degree of invariants can be read off the series, but also their composition in terms of different representations. % of the invariants can be read off from the series, e.g, degree-six invariant is built from three instances of $\vec{\Lambda}$ and three instances of (the traceless part of) $\Lambda_{ij}$.
Again, the denominator corresponds to the basic invariants $I_{i,j}$ \eqref{eq:quartic_invariants}.  % :
%\begin{align}
%	I_{1,1} & = \lamS, \hspace{1.5cm}
%I_{1,2}  = \trlam{}{},  \\ 
%I_{2,1} & =  \lamV \cdot \lamV, \hspace{1.3cm}
%I_{2,2}   = \trlam{}{2}, \\
%I_{3,1} & = \lamV \cdot \lamT \cdot \lamV,  \hspace{0.8cm}
%I_{3,2}  = \trlam{}{3}, \\ 
%I_{4,1} & = \lamV \cdot \lamT^2 \cdot \lamV. 
%\end{align}
From the numerator of \eqref{eq:hilbert_series_sc} one deduces that an invariant of order six \eqref{eq:quartic_invariants_six} should be built from three instances of $\lamV$ and three instances of (the traceless part of) $\lamT$. %there is an invariant of order 6
%\begin{align}
%	I_{6,1}& = \lamV \cdot\left[(\lamT \cdot \lamV) \times (\lamT^2 \cdot \lamV)\right],
%\end{align}
%which 
One can also prove the following relation expressing the square of the degree-six invariant in terms of the invariants from the denominator \eqref{eq:quartic_invariants}
\begin{align}
	I_{6,1}^2 & = 
	  - I_{4,1}^3
          + I_{2,1} I_{3,1} I_{3,2} I_{4,1}
	  + 2 \left( 
		  I_{1,2} I_{3,1} I_{4,1}^2 -  I_{1,2} I_{2,2} I_{2,1} I_{3,1} I_{4,1}
  	      \right)
          + I_{1,2}^3 I_{2,1} I_{3,1} I_{4,1}
	  \nonumber\\
	  & 
	  + \frac{1}{2} \left(
	  		I_{2,2} I_{3,1}^2 I_{4,1}
          + %\frac{1}{2} 
	  		I_{2,2} I_{2,1} I_{4,1}^2
          + %\frac{1}{2} 
	  		I_{1,2} I_{2,2}^2 I_{2,1}^2 I_{3,1}
          - %\frac{1}{2} 
	  		I_{1,2}^2 I_{2,1} I_{4,1}^2
          + %\frac{1}{2} 
	  		I_{1,2}^2 I_{2,2} I_{2,1}^2 I_{4,1}
  \right)
  \nonumber\\
	& 
  + \frac{1}{3} \left( 
	    I_{1,2} I_{2,2} I_{2,1}^3 I_{3,2}
          + %\frac{1}{3} 
	  		I_{1,2}^2 I_{2,1}^2 I_{3,1} I_{3,2}
          + %\frac{1}{3} 
	  		I_{1,2}^3 I_{3,1}^3
		\right) 
          + I_{1,2}^2 I_{2,2} I_{2,1} I_{3,1}^2
		\nonumber\\
		& - \frac{1}{3}\left(
            %\frac{1}{3} 
	  		I_{3,1}^3 I_{3,2}
          + %\frac{1}{3} 
	  		I_{2,2} I_{2,1}^2 I_{3,1} I_{3,2}
          + %\frac{1}{3} 
	  		I_{1,2} I_{2,1} I_{3,1}^2 I_{3,2}
          + %\frac{1}{3} 
	  		I_{1,2} I_{2,1}^2 I_{3,2} I_{4,1}
  		\right)
          - \frac{5}{12} I_{1,2}^4 I_{2,1} I_{3,1}^2 
\nonumber \\
& 
- \frac{1}{4} \left( I_{2,2}^2 I_{2,1} I_{3,1}^2
          + %\frac{1}{4} 
	  I_{1,2}^2 I_{2,2}^2 I_{2,1}^3
  \right)
  + \frac{1}{6} \left( I_{1,2}^4 I_{2,2} I_{2,1}^3
          + %\frac{1}{6} 
	  	I_{1,2}^5 I_{2,1}^2 I_{3,1}
          - %\frac{1}{6} 
	  	I_{1,2}^4 I_{2,1}^2 I_{4,1}
  \right)
  \nonumber \\
  & 
  - \frac{1}{9} \left( I_{2,1}^3 I_{3,2}^2
          + %\frac{1}{9} 
	  I_{1,2}^3 I_{2,1}^3 I_{3,2}
  \right)
          - \frac{1}{36} I_{1,2}^6 I_{2,1}^3
          - \frac{2}{3} I_{1,2}^3 I_{2,2} I_{2,1}^2 I_{3,1}
          - \frac{3}{2} I_{1,2}^2 I_{3,1}^2 I_{4,1}.
	  \label{eq:I61_squared}
\end{align}
As a consequence, only the sign of $I_{6,1}$ is important.

	For convenience we also present the Hilbert Series for the case, when a singlet $\mS$ and a triplet $\mV$ originating from the mass term are taken into account,
\begin{align}
	H(t) = \frac{1 + t^3 + 4 t^4 + 2 t^5 + 4 t^6 + t^7 + t^{10}}{
		\left(1-t\right)^3
		\left(1-t^2\right)^4
		\left(1-t^3\right)^3
		\left(1-t^4\right)
	}.
\end{align}
From the denominator one can immediately deduce the number of physical parameters ($p=11$) of the scalar potential of 2HDM.

\section{Useful identities}

\label{app:identities}
	Here we list useful identities that were utilized to obtain compact expressions for the beta-functions and to check the corresponding pole equations.

From the Cayley-Hamilton theorem \eqref{eq:CayleyHamilton} valid for a $3\times 3$ matrix $\Lambda = A + h\, B$  with $A$ and $B$ also being matrices, at order $h$ we have
\begin{align}
	A^2 B + A B A  + B A^2 & =
	\left[ \tr{B} \cdot  A^2 + \tr{A} \cdot \left(A B + B A\right) \right]
	\nonumber\\
	&
	+ \frac{1}{2} \left(\tr^2{A}- \tr{A}^2 \right) B 
	+ \left( \tr{A} \cdot \tr{B} - \tr{(AB)} \right) A
	\nonumber \\
	& - \left[ \frac{1}{2} \tr^2 A \cdot \tr B 
	- \frac{1}{2} \tr{A^2} \cdot \tr B 
- \tr A \cdot \tr{(AB)} + \tr{(A^2 B)} \right].
\label{eq:ABA_identity}
\end{align}
This identity can be used to prove that the degree-four tensor%$eliminate the sum $A^2 B + B A^2$ in favour of $A B A$ for a degree-four tensor ($A=\lamT$, $B=\lamV \otimes \lamV$) %. %, where ($A = \lamT$, $B = \lamV \otimes \lamV$

\begin{align}
\left[ 
		\lamT^2 \cdot \lamV \otimes  \lamV 
		+ 
		 \lamV \otimes \lamT^2 \cdot \lamV 
		\right]
& = 
		- \lamT \cdot \lamV \otimes \lamT \cdot \lamV 		
		+ \lamV^2 \cdot \lamT^2  
		+ \trlam{}{} \left( 
		\lamT \cdot \lamV \otimes  \lamV 
		+ 
		 \lamV \otimes \lamT \cdot \lamV 
		\right)
		\nonumber\\
& 
		- \frac{1}{2} \left( \trlam{2}{} - \trlam{}{2} \right)
		\lamV \otimes \lamV
		- \lamT \left( \trlam{}{} \cdot \lamV^2 - ( \lamV \cdot \lamT \cdot \lamV) \right)
		\nonumber\\
		& 
		+
		\left[
		    \frac{\lamV^2}{2} \left( \trlam{2}{} - \trlam{}{2} \right)
		  - \trlam{}{} \cdot ( \lamV \cdot \lamT \cdot \lamV ) 
		  + ( \lamV \cdot \lamT^2 \cdot \lamV )  
		\right],
\label{eq:order4_tensor_dep}
\end{align}
and the degree-five tensor %multiply by $\lamT$ (left) 
\begin{align}
\left[ 
	\lamT^2 \cdot \lamV \otimes \lamT \cdot \lamV 			
	+ 
		 \lamT \cdot \lamV \otimes \lamT^2 \cdot \lamV 
		\right]
& = 
		- 
%
		%\lamT^3 \cdot \lamV \otimes  \lamV 
% ->
%		\left(
%		\trlam{}{} \, \lamT^2 \cdot \lamV \otimes  \lamV  
%		- \frac{1}{2} \left( \trlam{2}{} - \trlam{}{2} \right)
%		\lamT \cdot \lamV \otimes  \lamV 
%		+ 
		\frac{1}{3!} \left( 
			\trlam{3}{} - 3 \trlam{}{} \trlam{}{2} + 2 \trlam{}{3}
%		\right)
	\right)
	\left( \lamV \otimes  \lamV  - \lamV^2\right)
	\nonumber\\
	&
		+ \trlam{}{} \left( 
%		\lamT^2 \cdot \lamV \otimes  \lamV 
		%+ 
		 \lamT\cdot \lamV \otimes \lamT \cdot \lamV 
		\right)
		%\nonumber\\
%& 
%		- \frac{1}{2} \left( \trlam{2}{} - \trlam{}{2} \right)
%		\lamT\cdot\lamV \otimes \lamV
		+ \lamT^2 \left( %\trlam{}{} \cdot \lamV^2 
		  \lamV \cdot \lamT \cdot \lamV \right)
		\nonumber\\
		& 
		-
		\lamT \left[
		%    \frac{\lamV^2}{2} \left( \trlam{2}{} - \trlam{}{2} \right)
		   \trlam{}{} \cdot ( \lamV \cdot \lamT \cdot \lamV ) 
		  - ( \lamV \cdot \lamT^2 \cdot \lamV )  
		\right]
\label{eq:order5_tensor_dep}
\end{align}
can be reduced to the structures given in eq.~\eqref{eq:betb_LT_gen}. Moreover, the order-6 matrix 
\begin{align}
	& 
\left[ 
		 \lamT^2 \cdot \lamV \otimes \lamT^2 \cdot \lamV 
		\right]
 = 
	%- \lamT^3 \cdot \lamV \otimes \lamT \cdot \lamV 				
%		- \trlam{}{} \left( \lamT^2 \cdot \lamV \otimes \lamT \cdot \lamV 			\right)
	 \frac{1}{2} \left( \trlam{2}{} - \trlam{}{2} \right)
	\left(\lamT \cdot \lamV \otimes \lamT \cdot \lamV\right)
	%- \frac{1}{3!} \left( \trlam{3}{} - 3 \trlam{}{} \trlam{}{2} + 2 \trlam{}{3}
	%	\right)
	%	\left( \lamV \otimes \lamT \cdot \lamV 	\right)
	\nonumber\\
	&
	- 
%
		%\lamT^3 \cdot \lamV \otimes  \lamV 
% ->
%		\left(
%		\trlam{}{} \, \lamT^2 \cdot \lamV \otimes  \lamV  
%		- \frac{1}{2} \left( \trlam{2}{} - \trlam{}{2} \right)
%		\lamT \cdot \lamV \otimes  \lamV 
%		+ 
		\frac{1}{3!} \left( 
			\trlam{3}{} - 3 \trlam{}{} \trlam{}{2} + 2 \trlam{}{3}
%		\right)
	\right)
	\left[ 
	  \lamT\cdot \lamV \otimes  \lamV  
	+ \lamV \otimes \lamT \cdot \lamV 	
	- \lamV^2 \cdot \lamT 
	-
		\left( %\trlam{}{} \cdot \lamV^2 
		  \lamV \cdot \lamT \cdot \lamV \right)
\right]
	\nonumber\\
	&
	%	+ \trlam{}{} \left( 
%		\lamT^2 \cdot \lamV \otimes  \lamV 
		%+ 
%		 \lamT^2 \cdot \lamV \otimes \lamT \cdot \lamV 
%		\right)
		%\nonumber\\
%& 
%		- \frac{1}{2} \left( \trlam{2}{} - \trlam{}{2} \right)
%		\lamT\cdot\lamV \otimes \lamV
		%+ \lamT^3 
		+
		%    \frac{\lamV^2}{2} \left( \trlam{2}{} - \trlam{}{2} \right)
	%	   \trlam{}{} \cdot ( \lamV \cdot \lamT \cdot \lamV ) 
		    \left( \lamV \cdot \lamT^2 \cdot \lamV \right)  
		 \lamT^2 
		-
	%		\trlam{}{} \lamT^2 
		 \frac{1}{2} \left( \trlam{2}{} - \trlam{}{2} \right) %\lamT 
%		+ \frac{1}{3!} \left( 
%			\trlam{3}{} - 3 \trlam{}{} \trlam{}{2} + 2 \trlam{}{3}
%		\right)
		\left( %\trlam{}{} \cdot \lamV^2 
		  \lamV \cdot \lamT \cdot \lamV \right)
		  \lamT
		 .
\label{eq:order6_tensor_dep}
\end{align}
is also reducible.

In addition, from \eqref{eq:CayleyHamilton} the following identities that were utilized in due course of our calculations can be derived
\begin{align}
	\lamT^2 & - \trlam{}{} \lamT 
		+ \frac{1}{2} \left( \trlam{2}{} - \trlam{}{2} \right)
		- \det (\lamT) \cdot \lamT^{-1} = 0 \Rightarrow	
		\label{eq:CayleyHamilton2}
		\\
\epsilon_{\gamma i j} \lamT_{i \alpha} \lamT_{j \beta} & =
		\epsilon_{k \alpha \beta} \frac{\partial \det(\lamT) }{\partial \lamT_{k \gamma}} 
		= \epsilon_{k \alpha \beta} \left[ \det(\lamT) \lamT^{-1}_{\gamma k} \right]
		\Rightarrow
		\\
\epsilon_{\gamma i j} \lamT_{i \alpha} \lamT_{j \beta} & =
		\epsilon_{k \alpha \beta} \left[ \lamT^2_{\gamma k}
			- \trlam{}{}\cdot \lamT_{\gamma k} 
			+ \frac{1}{2} \delta_{\gamma k} 
			\left( \trlam{2}{} - \trlam{}{2} \right)
		\right].
	\label{eq:CayleyHamiltonEps}
\end{align}

Finally, it is interesting to note the relation valid in 3d for arbitrary vectors $\vec{a}$ and $\vec{b}$ and arbitrary symmetric matrix $C$ 
\begin{align}
	\vec{a} \times \vec{b} \cdot \tr{C}  
	= C \cdot (\vec{a} \times \vec{b}) +
	(C \cdot \vec{a}) \times \vec{b} +
	\vec{a} \times (C \cdot \vec{b}).
\end{align}
The relation was used to simplify the result for the mass-term RGE.

\section{\label{app:invariants_rge}Scale dependence of the invariants}
The RG functions for reparametrization invariants %from eq.~\eqref{eq:quartic_invariants} 
are defined as
\begin{align}
	\frac{d I_{i,j}}{d t} & = \sum_{l=1}^\infty 
	h^l \beta^{(l)}_{i,j}, \quad t=\ln \mu^2, \quad h = \frac{1}{16\pi^2}.
\end{align}
We present here the expressions for $\beta^{(l)}_{i,j}$ up to the two-loop order\footnote{The three-loop contribution can be found online as ancillary files of the arXiv version of the paper.}. The beta-functions for degree-one invariants have the form:
\begin{align}
2\beta^{(1)}_{1,1} & = 
    8I_{1,1}^2
    +2\left[I_{1,1} I_{1,2}+I_{2,2}\right]
    +12I_{2,1},
 \label{eq:beta_I_11_1}
 \\
2\beta^{(2)}_{1,1} & = 
    \frac{5}{2}I_{1,1} I_{1,2}^2
    -132I_{1,1} I_{2,1}
    -98I_{3,1}
    -\frac{57}{2}I_{1,1}^3
    \nonumber \\
    &
    -27I_{1,1} I_{2,2}
    -16I_{3,2}
    -11I_{1,1}^2 I_{1,2}
    -4I_{1,2} I_{2,1}
    +2I_{1,2} I_{2,2},
    \label{eq:beta_I_11_2}
%\end{align}
\\
%\begin{align}
2\beta^{(1)}_{1,2} & = 
    8I_{2,2}
    -2I_{1,2}^2
    +6I_{1,1} I_{1,2}
    +12I_{2,1},
 \label{eq:beta_I_12_1}
\\
2\beta^{(2)}_{1,2} & = 
	-122I_{3,1}
    -102I_{1,1} I_{2,1}
    -48I_{1,1} I_{2,2}
    -\frac{61}{2}I_{1,1}^2 I_{1,2}
    \nonumber \\
    &
    -16I_{3,2}
    -10I_{1,2} I_{2,1}
    -\frac{3}{2}I_{1,2}^3
    +7I_{1,1} I_{1,2}^2
    +11I_{1,2} I_{2,2}.
 \label{eq:beta_I_12_2}
\end{align}
The scale dependence of degree-two and degree-three invariants are given by
{\allowdisplaybreaks
\begin{align}
2\beta^{(1)}_{2,1} & = 
    I_{1,1} I_{2,1}+I_{3,1},
 \label{eq:beta_I_21_1}
\\
2\beta^{(2)}_{2,1} & = 
    I_{1,2}^2 I_{2,1}
    -204I_{1,1} I_{3,1}
    -156I_{2,1}^2
    -127I_{1,1}^2 I_{2,1},
    \nonumber \\
    &
    -122I_{4,1}
    -14I_{2,1} I_{2,2}
    -10I_{1,1} I_{1,2} I_{2,1}
    +8I_{1,2} I_{3,1},
 \label{eq:beta_I_21_2}
%\end{align}
\\
%\begin{align}
2\beta^{(1)}_{2,2} & = 
    16I_{3,2}
    -4I_{1,2} I_{2,2}
    +12I_{1,1} I_{2,2}
    +24I_{3,1},
 \label{eq:beta_I_22_1}
\\
2\beta^{(2)}_{2,2} & = 
    6I_{2,2}^2
    -244I_{4,1}
    -204I_{1,1} I_{3,1}
    -96I_{1,1} I_{3,2}
    -61I_{1,1}^2 I_{2,2}
    \nonumber \\
    &
    -\frac{128}{3}I_{1,2} I_{3,2}
    -28I_{2,1} I_{2,2}
    -\frac{16}{3}I_{1,2}^4
    +8I_{1,2} I_{3,1}
    +14I_{1,1} I_{1,2} I_{2,2}
    +29I_{1,2}^2 I_{2,2},
 \label{eq:beta_I_22_2}
%\end{align}
\\
%\begin{align}
2\beta^{(1)}_{3,1} & = 
     12I_{2,1}^2
    -2I_{1,2} I_{3,1}
    +30I_{1,1} I_{3,1}
    +32I_{4,1}
 \label{eq:beta_I_31_1}
 \\
2\beta^{(2)}_{3,1} & = 
     4I_{1,2} I_{2,1}^2
    -292I_{2,1} I_{3,1}
    -252I_{1,1} I_{4,1}
    -\frac{315}{2}I_{1,1}^2 I_{3,1},
    \nonumber \\
    &
    -130I_{1,2} I_{4,1}
    -102I_{1,1} I_{2,1}^2
    -72I_{2,2} I_{3,1}
    -46I_{2,1} I_{3,2}
    \nonumber \\
    &
    -23I_{1,2}^3 I_{2,1}
    -3I_{1,1} I_{1,2} I_{3,1}
    +\frac{137}{2}I_{1,2}^2 I_{3,1}
    +69I_{1,2} I_{2,1} I_{2,2},
 \label{eq:beta_I_31_2}
%\end{align}
\\
%\begin{align}
2\beta^{(1)}_{3,2} & = 
     12I_{2,2}^2
    -24I_{1,2}^2 I_{2,2}
    +4I_{1,2}^4
    +18I_{1,1} I_{3,2}
    +26I_{1,2} I_{3,2}
    +36I_{4,1},
 \label{eq:beta_I_32_1}
\\
2\beta^{(2)}_{3,2} & = 
     40I_{1,2}^3 I_{2,2}
    -354I_{1,2} I_{4,1}
    -306I_{1,1} I_{4,1}
    -171I_{1,1} I_{1,2} I_{3,2}
    \nonumber \\
    &
    -164I_{2,1} I_{3,2}
    -\frac{183}{2}I_{1,1}^2 I_{3,2}
    -72I_{1,1} I_{2,2}^2
    -61I_{1,2}^3 I_{2,1}
    \nonumber \\
    &
    -\frac{89}{2}I_{1,2}^2 I_{3,2}
    -24I_{1,1} I_{1,2}^4
    -8I_{1,2}^5
    -7I_{2,2} I_{3,2}
    \nonumber \\
    &
    +144I_{1,1} I_{1,2}^2 I_{2,2}
    +183\left[I_{1,2} I_{2,1} I_{2,2}+I_{1,2}^2 I_{3,1}-I_{2,2} I_{3,1}\right].
 \label{eq:beta_I_32_2}
\end{align}
}
The degree-four invariant depends on the scale as   
\begin{align}
2\beta^{(1)}_{4,1} & = 
     24I_{2,1} I_{3,1}
    +\frac{20}{3}I_{1,2}^3 I_{2,1}
    +\frac{40}{3}I_{2,1} I_{3,2}
    +36\left[I_{1,1} I_{4,1}+I_{1,2} I_{4,1}\right]
    \nonumber \\
    &
    -20\left[I_{1,2} I_{2,1} I_{2,2}+(I_{1,2}^2-I_{2,2} )I_{3,1}\right],
 \label{eq:beta_I_41_1}
\\
2\beta^{(2)}_{4,1} & = 
     4I_{1,2} I_{2,2} I_{3,1}
    -306I_{2,1} I_{4,1}
    -296I_{1,1} I_{1,2} I_{4,1}
    -204I_{1,1} I_{2,1} I_{3,1}
    -188I_{1,1}^2 I_{4,1}
    \nonumber \\
    &
    -122I_{3,1}^2
    -100I_{1,1} I_{2,1} I_{3,2}
    -71I_{1,2}^2 I_{4,1}
    -69I_{2,2} I_{4,1}
    -\frac{154}{3}I_{3,1} I_{3,2}
    -50I_{1,1} I_{1,2}^3 I_{2,1}
    \nonumber \\
    &
    -\frac{146}{3}I_{1,2} I_{2,1} I_{3,2}
    -\frac{73}{3}I_{1,2}^4 I_{2,1}
    +8I_{1,2} I_{2,1} I_{3,1}
    +\frac{142}{3}I_{1,2}^3 I_{3,1}
    \nonumber \\
    &
    +73I_{1,2}^2 I_{2,1} I_{2,2}
    +150\left[I_{1,1} I_{1,2} I_{2,1} I_{2,2}+I_{1,1} I_{1,2}^2 I_{3,1}-I_{1,1} I_{2,2} I_{3,1}\right].
 \label{eq:beta_I_41_2}
\end{align}
One can see that the beta-functions for \eqref{eq:quartic_invariants} do not depend on the degree-six invariant $I_{6,1}$, while the latter has the following beta-function 
\begin{align}
	2\beta^{(1)}_{6,1} & = 
	I_{6,1} 
	\left(22I_{1,2} +54 I_{1,1}\right),
	\label{eq:beta_I_61_1}
\\
	2\beta^{(2)}_{6,1} & = 
	-I_{6,1} 
	\left(
     398I_{2,1} 
    +282I_{1,1}^2 
    +192I_{1,1} I_{1,2} 
    +73I_{2,2} 
    +7I_{1,2}^2 
    \right).
	\label{eq:beta_I_61_2}
\end{align}
It is worth noting that we have checked that the three-loop RGE respect the relation \eqref{eq:I61_squared}. 

The RG functions for the invariants involving mass parameters have the form:
{\allowdisplaybreaks
\begin{align}
2\beta^{(1)}_{0,1} & = 
    6\,I_{1,3}
    +\left(I_{1,2}+5I_{1,1}\right)\,I_{0,1},
 \label{eq:beta_I_01_1}
\\
2\beta^{(2)}_{0,1} & = 
    \left(\frac{5}{4}I_{1,2}^2-15I_{2,1}-\frac{15}{2}I_{2,2}-\frac{25}{4}I_{1,1}^2-\frac{5}{2}I_{1,1} I_{1,2}\right)\,I_{0,1}
    -15\,I_{2,3}
    -15I_{1,1}\,I_{1,3},
 \label{eq:beta_I_01_2}
%\end{align}
\\
%\begin{align}
2\beta^{(1)}_{1,3} & = 
    18\,I_{2,3}
    -\left(I_{1,2}-13I_{1,1}\right)\,I_{1,3}
    +6I_{2,1}\,I_{0,1},
 \label{eq:beta_I_13_1}
\\
2\beta^{(2)}_{3,1} & = 
    \left(4I_{1,2}-114I_{1,1}\right)\,I_{2,3}
    -67\,I_{3,3}
    -15\left[I_{1,1} I_{2,1}+I_{3,1}\right]\,I_{0,1}
    \nonumber \\
    &
    - \left(93I_{2,1}+\frac{263}{4}I_{1,1}^2+\frac{7}{2}I_{1,1} I_{1,2}+\frac{5}{2}I_{2,2}+\frac{1}{4}I_{1,2}^2\right)\,I_{1,3},
 \label{eq:beta_I_13_2}
%\end{align}
\\
%\begin{align}
2\beta^{(1)}_{2,3} & = 
    \left(19I_{1,1}-3I_{1,2}\right)\,I_{2,3}
    +26\,I_{3,3}
    +6I_{3,1}\,I_{0,1}
    +12I_{2,1}\,I_{1,3},
 \label{eq:beta_I_23_1}
\\
2\beta^{(2)}_{2,3} & = 
    \left(\frac{7}{2}I_{1,1} I_{1,2}+\frac{159}{4}I_{1,2}^2-150I_{2,1}-\frac{385}{4}I_{1,1}^2-33I_{2,2}\right)\,I_{2,3}
    \nonumber \\
    &
    +\left(4I_{1,2} I_{2,1}+\frac{83}{2}I_{1,2} I_{2,2}-102I_{1,1} I_{2,1}-79I_{3,1}-\frac{83}{3}I_{3,2}-\frac{83}{6}I_{1,2}^3\right)\,I_{1,3}
    \nonumber \\
    &
    -\left(162I_{1,1}+79I_{1,2}\right)\,I_{3,3}
    -15\left[I_{1,1} I_{3,1}+I_{4,1}\right]\,I_{0,1},
 \label{eq:beta_I_23_2}
%\end{align}
\\
%\begin{align}
2\beta^{(1)}_{3,3} & = 
    6I_{4,1}\,I_{0,1}
    +\left(12I_{2,1}+17\left[I_{2,2}-I_{1,2}^2\right]\right)\,I_{2,3}
    \nonumber \\
    &
    +\left(25I_{1,1}+29I_{1,2}\right)\,I_{3,3}
    +\left(\frac{17}{3}I_{1,2}^3+\frac{34}{3}I_{3,2}+12I_{3,1}-17I_{1,2} I_{2,2}\right)\,I_{1,3},
 \label{eq:beta_I_33_1}
\\
2\beta^{(2)}_{3,3} & = 
\left(\frac{15}{2}\left[I_{1,2} I_{2,1} I_{2,2}+(I_{1,2}^2 -I_{2,2}) I_{3,1}\right]-5I_{2,1} I_{3,2}
-15I_{4,1}\left[I_{1,1} +I_{1,2} \right]
\right. \nonumber \\
&  \left.
-\frac{5}{2}I_{1,2}^3 I_{2,1}
\right)\,I_{0,1}
%\nonumber \\
%& 
- \left(\frac{399}{2}I_{1,1} I_{1,2}+164I_{2,1}+\frac{507}{4}I_{1,1}^2+\frac{195}{4}I_{1,2}^2+30I_{2,2}\right)\,I_{3,3}  
\nonumber \\
&
+\left(\vphantom{\frac{3}{2}}
2I_{1,2} I_{2,2}+4I_{1,2} I_{2,1}+31I_{1,2}^3+105 I_{1,2}\left[I_{1,2}^2- I_{2,2}\right]
-122I_{3,1}-102I_{1,1} I_{2,1}
\right. \nonumber \\
& \left.
\vphantom{\frac{3}{2}}
-33I_{3,2}\right)\,I_{2,3}
%\nonumber \\
%&
    +\left(4I_{1,2} I_{3,1}+\frac{95}{2}I_{1,2}^2 I_{2,2}+105I_{1,1} I_{1,2} I_{2,2}-102I_{1,1} I_{3,1}
    \right. \nonumber \\
    & \phantom{+} \left.
    -79I_{4,1}-70I_{1,1} I_{3,2}-35I_{1,1} I_{1,2}^3-\frac{95}{3}I_{1,2} I_{3,2}-\frac{95}{6}I_{1,2}^4\right)\,I_{1,3}
 \label{eq:beta_I_33_2}.
\end{align}
}

	Finally, let us introduce a new set of invariants $\tilde I_{i,j}$ (c.f., ref.\cite{Ivanov:2005hg}),
	in which $\Lambda$  is replaced by its traceless part $\tilde \lamT  = \lamT - \frac{1}{3} \trlam{}{}$ in all $I_{i,j}$ %\eqref{eq:quartic_invariants} and \eqref{eq:quartic_invariants_six} 
	but $I_{1,2}$. Obviously, $I_{0,1}$, $I_{1,1}$, $I_{1,3}$, and $I_{2,1}$ do not involve $\Lambda$, while other invariants change as 
\begin{align}
	\tilde I_{2,2} & \equiv \tr{\tilde \lamT^2} = 
	I_{2,2} - \frac{1}{3} I_{1,2}^2,	
	\nonumber \\
	\tilde I_{2,3} &  \equiv \lamV \cdot \tilde \lamT \cdot \mV
= I_{2,3} - \frac{1}{3} I_{1,2} I_{1,3},\
	\nonumber \\
	\tilde I_{3,1} &  \equiv \lamV\cdot\tilde\lamT \cdot \lamV = I_{3,1} - \frac{1}{3} I_{1,2} I_{2,1},
	\nonumber\\
	\tilde I_{3,2} &  \equiv \tr \tilde \lamT^3 =  I_{3,2} - I_{1,2} I_{2,2} + \frac{2}{9} I_{1,2}^3, 
\nonumber\\ 
\tilde I_{3,3} & \equiv \lamV \cdot \tilde \lamT^2 \cdot \mV = I_{3,3} - \frac{2}{3} I_{1,2} I_{2,3} + \frac{1}{9} I_{1,2}^2 I_{1,3},
%\quad
\nonumber\\
\tilde I_{4,1}  & = \lamV \cdot \tilde \lamT^2 \cdot \lamV = I_{4,1} - \frac{2}{3} I_{1,2} I_{3,1} + \frac{1}{9} I_{1,2}^2 I_{2,1},
\nonumber \\
\tilde I_{6,1} & \equiv \lamV \cdot \left[
	( \tilde \lamT \cdot \lamV )
	\times 
	( \tilde \lamT^2 \cdot \lamV )
\right] = I_{6,1}.
\label{eq:invariants_set_2}
\end{align}
By simple algebra one can rewrite all the results in terms of \eqref{eq:invariants_set_2}.

%\section*{References}

%\bibliography{2hdm_rge}

\providecommand{\href}[2]{#2}\begingroup\raggedright\endgroup

\end{document}